\documentclass[aps,prc,twocolumn,10pt, superscriptaddress, showpacs, floatfix]{revtex4-1}
\usepackage{amssymb,epsfig}

\graphicspath{{figures/}}
\newcommand{\be}{\begin{equation}}
\newcommand{\ee}{\end{equation}}
\newcommand{\bea}{\begin{eqnarray}}
\newcommand{\eea}{\end{eqnarray}}

\begin{document}



\title{Soft modes in the proton-neutron pairing channel as precursors of deuteron condensate in N=Z nuclei}
\author{Elena Litvinova}
\affiliation{Department of Physics, Western Michigan University, Kalamazoo, MI 49008, USA}
\affiliation{National Superconducting Cyclotron Laboratory, Michigan State University, East Lansing, MI 48824, USA}
\author{Caroline Robin}
\affiliation{Institute for Nuclear Theory, University of Washington, Seattle, WA 98195, USA}
\affiliation{JINA-CEE, Michigan State University, East Lansing, MI 48824, USA}
\author{Irina A. Egorova}
\affiliation{Department of Physics, Western Michigan University, Kalamazoo, MI 49008, USA}
\affiliation{Bogoliubov Laboratory of Theoretical Physics, JINR, Dubna, 141980, Russia}
\date{\today}

\begin{abstract}
Relativistic nuclear response theory is formulated for the proton-neutron pairing, or deuteron transfer, channel.  The approach is based on the meson-nucleon Lagrangian of Quantum Hadrodynamics (QHD) and advances the relativistic field theory to connect consistently the high-energy scale of heavy mesons, the medium-energy range of the pion and the low-energy domain of emergent collective vibrations (phonons) in a parameter-free way. Mesons and phonons build up the in-medium nucleon-nucleon interaction in spin-isospin transfer channels, in particular, the phonon-exchange part takes care of the leading-order retardation effects. In this framework, we explore 
$J^{\pi} = 0^+$ and $J^{\pi} = 1^+$ channels of the nuclear response to the proton-neutron pair removal and addition in $^{56}$Ni and $^{100}$Sn with a special focus on the lowest (soft) modes as precursors of deuteron condensate and candidates for being the mediators of the proton-neutron pairing interaction. 

\end{abstract}


\maketitle


\section{Introduction}
\label{sec:intro}


The complex nature of nuclear forces remains at the frontiers of fundamental physics and continuously raises questions regarding
the behavior of atomic nuclei under various conditions and their response to external probes. One of those questions is the impact of superfluidity on nuclear static and dynamic characteristics \cite{RingSchuck1980}.
%
%
 In many nuclear structure approaches this type of nucleon-nucleon interaction is conventionally separated from other correlations because of its similarities to the electron-electron correlations in metals. Already in the end of the 1950-s it has become clear that a large corpus of experimental data can be  explained by the presence of superfluidity in nuclei \cite{BohrMottelsonPines1958}. Since then, to include the pairing correlations between nucleons of the same isospin and coupled to zero angular momentum by the Bardeen-Cooper-Schieffer (BCS) or Bogoliubov's formalism has become a standard procedure for nuclear models, although more extended approaches have also been developed \cite{50BCS}. In particular, the influence of collective vibrations on such pairing was investigated 
\cite{BarrancoBrogliaGoriEtAl1999,AvdeenkovKamerdzhiev1999,BarrancoBortignonBrogliaEtAl2005,IdiniPotelBarrancoEtAl2015} and found significant in being responsible for up to 
half of the total pairing gap. Dynamical aspects of pairing can be quantified by the nuclear response to the pair transfer, accessible by the particle-particle random phase approximation (pp-RPA) \cite{BesBroglia1966,RipkaPadjen1969} and its extensions including retardation mechanisms, such as the particle-vibration coupling (PVC) \cite{HahneHeissEngelbrecht1977}. 

The common viewpoint is that, because of isospin symmetry, it is also natural to expect superfluid pairing correlations (pairing in the following) between protons and neutrons in nuclei. A direct evidence for the presence of proton-neutron pairing (pn-pairing) is the fact that the ground states of the most odd-odd $N=Z$ nuclei with mass $A <$ 40 have isospin T=0 and spin J$>$0, while the situation changes for their counterparts with $A >$ 40, where the ground states have T=1 and J=0 with the only known exception of $^{58}$Cu \cite{EngelLangankeVogel1996,LangankeMartinez-Pinedo2013}. 
Theoretical studies reported in Refs. \cite{SatulaWyss1997,Goodman1999,BertschLuo2010,GezerlisBertschLuo2011a,Yoshida2014} discuss the possibility of T=0 pairing condensate in heavy N$\sim$Z nuclei as a consequence of the attractive proton-neutron interaction in the $^3S_1$ channel. 

A recent comprehensive analysis of the existing experimental and theoretical results obtained by various groups \cite{Frauendorf201424} concludes that there are evident signatures for (T=1, J=0) pn-pairing, in particular, "the binding energies of the even-A nuclides around the double magic nuclei $^{40}$Ca and $^{56}$Ni organize into a pattern of soft (T=1, J=0) pair vibrational excitations, which represents the
precursor of the isovector pair condensate appearing further in the open shell." In contrast, no strong collectivity was found in the (T=0, J=1) pn-pairing channel, and  most of the experimental data speaks against the existence of a (T=0, J=1) deuteron-like pair condensate.
The authors suggest that the isoscalar pn-pairing can possibly only occur in a form of phonons in the $30 < A < 100$ mass region, but not likely as a condensate. However, theoretical mean-field calculations for $A > 200$ show that the weakening of the spin-orbit interaction may favor (T=0, J=1) ground state solutions \cite{Frauendorf201424}. Regarding infinite nuclear systems, recent studies of the gap equation with medium polarization effects \cite{ZhangCaoLombardoEtAl2016} showed that the superfluid spin-triplet phase disappears in asymmetric nuclear matter.

Summarizing the present understanding of the proton-neutron pairing, one can conclude that both singlet and triplet as well as, in principle, higher-spin pairing may occur in nuclei although not inevitably in the form of a condensate, but rather as a phonon exchange. In this work, based on our previous developments, we will try to identify possible underlying dynamical mechanisms of pn-pairing and to describe it consistently in the meson-exchange theory.
%
In Ref. \cite{RobinLitvinova2016} we have developed an approach to nuclear spin-isospin response which is based on the relativistic QHD Lagrangian and includes consistently meson exchange, superfluid pairing between like particles and the exchange of collective vibrations. Compared to the proton-neutron relativistic QRPA (pn-RQRPA) \cite{NiksicMarketinVretenarEtAl2005}, our approach called proton-neutron relativistic quasiparticle time blocking approximation (pn-RQTBA) has an additional quasiparticle-vibration coupling (QVC) mechanism of the nucleon-nucleon interaction which takes care of the retardation effects. Within this approach, GTR and beta-decay half-lives in the chain of nickel isotopes $^{68-78}$Ni were calculated without adopting the phenomenological pn-pairing.        
Comparing these results of the pn-RQTBA calculations to that of the relativistic and non-relativistic pn-QRPA with the phenomenological proton-neutron pairing \cite{EngelBenderDobaczewskiEtAl1999,NiksicMarketinVretenarEtAl2005}, one can notice that both QVC and pn-pairing shift and/or redistribute the Gamow-Teller strength to lower energies. Indeed, our analysis of the diagrammatic structure of the pn-RQTBA 
response function shows that, due to the presence of abnormal Gor'kov propagators, it 
contains terms with the phonon exchange between the proton and neutron particle-particle components, which can be interpreted as dynamical analogs of the pn-pairing. Such terms occur in the superfluid systems where quasiparticles are linear superpositions of particles and holes. However, the first version of the pn-RQTBA \cite{RobinLitvinova2016} contains only normal phonons of natural parities, among which the spins J $= 2-6$ provide the major contribution while the phenomenological pn-pairing is associated with (T=0, J$>$0) and (T=1, J=0) interactions. Thus, if this type of pairing occurs in the form of phonons, they should be the proton-neutron pairing phonons with these quantum numbers. 

The microscopic nature of the isospin-flip modes and their influence on the nuclear shell structure and on the spin-isospin response was discussed in Refs. \cite{Litvinova2016,RobinLitvinova2017}, respectively. It has been shown that coupling of the single-particle states to these modes induces further redistribution of the strength functions, compared to the case of coupling to only normal phonons.
%
%
However, so far only the particle-hole nuclear response functions were investigated in the relativistic framework. In this work we consider particle number violating excitation modes, or proton-neutron pair vibrations and present the relativistic time blocking approximation (RTBA) in the proton-neutron particle-particle (pn-pairing) channel. In the following it will be abbreviated as pn-pp-RTBA to be distinguished from the RTBA in the neutral particle-hole channel
developed originally in Ref. \cite{LitvinovaRingTselyaev2007} and from the proton-neutron particle-hole RTBA (pn-RTBA) of Refs. \cite{MarketinLitvinovaVretenarEtAl2012,LitvinovaBrownFangEtAl2014}. 

The method is based on the relativistic meson-exchange nuclear QHD Lagrangian and extends the covariant response theory by effects of retardation induced by a strongly-correlated medium. The retardation, or time dependence of the in-medium meson-exchange interaction, is neglected in the covariant density functional theory limited by the Hartree(Fock) approximation \cite{Ring1996,VretenarAfanasjevLalazissisEtAl2005,LiangVanGiaiMengEtAl2008} and in the approaches on the level of two-quasiparticle configurations, such as (relativistic) QRPA \cite{PaarRingNiksicEtAl2003,NiksicMarketinVretenarEtAl2005}. The time dependence is restored in the pn-pp-RTBA in an approximate way, taking into account the most important  (resonant) effects of temporal non-localities, essential at the relevant excitation energies ($\sim$ 0-50 MeV).  In the original version of RTBA \cite{LitvinovaRingTselyaev2007} they are modeled by coupling of particle-hole pairs to collective vibrations within the "particle-hole plus phonon" (ph$\otimes$phonon) coupling scheme, and the extended versions include like-particle superfluid pairing \cite{LitvinovaRingTselyaev2008}, phonon-phonon coupling \cite{LitvinovaRingTselyaev2010,LitvinovaRingTselyaev2013} and 
"two-quasiparticles plus N phonons" (2q$\otimes$Nphonon) \cite{Litvinova2015} configurations. The method was applied successfully to multipole spectra of various closed and open-shell medium-mass nuclei \cite{LitvinovaRingTselyaev2008,LitvinovaRingTselyaevEtAl2009,MassarczykSchwengnerDoenauEtAl2012,EgorovaLitvinova2016}. In particular, it has described well the observed isospin splitting of pygmy dipole resonance \cite{EndresLitvinovaSavranEtAl2010,EndresSavranButlerEtAl2012,LanzaVitturiLitvinovaEtAl2014}, isoscalar dipole modes \cite{PellegriBraccoCrespiEtAl2014,KrzysiekKmiecikMajEtAl2016}, and stellar reaction rates of the r-process nucleosynthesis \cite{LitvinovaLoensLangankeEtAl2009}. Current developments are focused on the generalized RQTBA \cite{Litvinova2015} which is shown to be similar to the equation of motion method for time-dependent density matrices \cite{SchuckTohyama2016a} in the sector describing the time evolution of two-body propagators \cite{LitvinovaSchuck2017}.

Thus, this work continues a series of extensions of the R(Q)TBA to spin-isospin excitations \cite{MarketinLitvinovaVretenarEtAl2012,LitvinovaBrownFangEtAl2014,RobinLitvinova2016}. The formalism of the pn-pp-RTBA is presented in Section \ref{sec:thmod}, Section \ref{sec:disc} discusses numerical details and results for $^{56}$Ni and $^{100}$Sn, and Section \ref{summary} gives conclusions and outlook.


\section{Formalism}
\label{sec:thmod}



The linear response of a nucleus to the proton-neutron pair (deuteron) transfer (removal or addition) can be described by the Bethe-Salpeter equation (BSE) projected onto the proton-neutron particle-particle channel \cite{RingSchuck1980}. It can be conveniently  formulated in the basis of states of Dirac-Hartree mean-field solutions $\{p_i\},\{n_i\}$ \cite{Ring1996}, where the indices $p_i, n_i$ run over the complete sets of the proton and neutron single-particle quantum numbers including states in the Dirac sea. In practice, this basis is generated by a self-consistent solution of the relativistic mean-field (RMF) problem \cite{Ring1996,VretenarAfanasjevLalazissisEtAl2005}. In a complete response theory, both static and dynamic terms of the in-medium nucleon-nucleon interaction enter the integral part of the BSE, which takes the form of a Dyson equation in the energy domain \cite{DukelskyRoepkeSchuck1998}. This is also the case for the resonant time blocking approximation \cite{Tselyaev1989,LitvinovaTselyaev2007,LitvinovaRingTselyaev2007}, in which the BSE for the response function $R(\omega)$ of the proton-neutron pair transfer takes the following form:
\begin{eqnarray}
R_{p_{1}n_{2},p_{3}n_{4}}^{\eta\eta^{\prime}}(\omega) = \tilde{R}_{p_{1}n_{2}%
}^{(0)\eta}(\omega)\delta_{p_{1}p_{3}}\delta_{n_{2}n_{4}}\delta_{\eta
\eta^{\prime}} +\nonumber \\
+ \tilde{R}_{p_{1}n_{2}}^{(0)\eta}(\omega)\sum\limits_{p_{5}%
n_{6} \eta^{\prime\prime}}{\bar{W}}_{p_{1}n_{2}%
,p_{5}n_{6}}^{\eta\eta^{\prime\prime}}(\omega)R_{p_{5}n_{6},p_{3}n_{4}}%
^{\eta^{\prime\prime}\eta^{\prime}}(\omega),
\label{respdir}
\end{eqnarray}
which is an analog of the BSE projected onto the particle-hole channel \cite{Tselyaev2007, LitvinovaTselyaev2007,MarketinLitvinovaVretenarEtAl2012,LitvinovaBrownFangEtAl2014}. Here the upper indices $\eta = \pm 1$ denote particle-particle (pp) and hole-hole (hh) nature of the respective proton-neutron pairs and $\tilde{R}^{(0)\eta}(\omega)$ is the propagator of an uncorrelated proton-neutron pair in the mean field
\begin{equation}
\tilde{R}_{p_{1}n_{2}}^{(0)\eta}(\omega) = \frac{\eta}{\omega - \varepsilon_{p_1} - \varepsilon_{n_2} + i\eta\delta}, \ \ \ \ \ \ \delta \to +0
\label{R0}
\end{equation}
 between acts of interaction with the following amplitude:
\begin{eqnarray}
{\bar{W}}_{p_{1}n_{2},p_{3}n_{4}}^{\eta\eta^{\prime}}(\omega)&=&\tilde{V}%
_{p_{1}n_{2},p_{3}n_{4}}^{\eta\eta^{\prime}} \nonumber \\
&&+\Bigl(\Phi_{p_{1}n_{2},p_{3}%
n_{4}}^{\eta}(\omega)-\Phi_{p_{1}n_{2},p_{3}n_{4}}^{\eta}(0)\Bigr)\delta
_{\eta\eta^{\prime}}.\nonumber \\
&& 
\label{W-omega}%
\end{eqnarray}
In this work, the operator $\tilde V$ is the meson-exchange interaction with the parameter set NL3$^*$  \cite{LalazissisKaratzikosFossionEtAl2009}
including a non-linear self-coupling of the scalar $\sigma$-meson:
\begin{equation}
{\tilde V} = {\tilde V}^{nl}_{\sigma} + {\tilde V}_{\omega} + {\tilde V}_{\rho} + {\tilde V}_{\pi} + {\tilde V}_{\delta\pi} + {\tilde V}_e, 
\label{mexch}
\end{equation}
where the time dependence is neglected in meson and photon propagators. In the proton-neutron response, however, only $\rho$-meson and pion contribute to $\tilde V$:
\begin{eqnarray}
{\tilde V}_{\rho}(1,2) =
g_{\rho}^2(\vec\tau\gamma_0\gamma^{\mu})_1(\vec\tau\gamma_0\gamma_{\mu})_2
D_{\rho}({\bf r}_1,{\bf r}_2) \nonumber\\
{\tilde V}_{\pi}(1,2) = -
\Bigl(\frac{f_{\pi}}{m_{\pi}}\Bigr)^{2}({\vec\tau}\ {\bf\Sigma}{\bf\nabla})_1
({\vec\tau}\ {\bf\Sigma}{\bf\nabla})_2D_{\pi}({\bf r}_1,{\bf r}_2),
\label{inter}
\end{eqnarray}
where $D_{\rho}$ and $D_{\pi}$ are the meson propagators, $g_{\rho}$ is the renormalized $\rho$-meson coupling constant of the NL3$^*$ parameter set and $f_{\pi}$ is the free-space pion-nucleon coupling.
The operator ${\bf\Sigma}$
is the generalized Pauli matrix \cite{PaarNiksicVretenarEtAl2004a}. The Landau-Migdal
term $V_{\delta\pi}$ is the part of the nucleon-nucleon
interaction providing its correct short-range behavior:
\begin{equation}
{\tilde V}_{\delta\pi}(1,2) =
g^{\prime}\Bigl(\frac{f_{\pi}}{m_{\pi}}\Bigr)^2({{\vec\tau}\ \bf\Sigma})_1({{\vec\tau}\ \bf\Sigma})_2
\delta({\bf r}_1 - {\bf r}_2),
\label{LM}
\end{equation}
where the parameter $g^{\prime}$ = 0.6 is used in the relativistic approaches without explicit Fock term while it has another fixed value in those treating the Fock term explicitly \cite{LiangVanGiaiMengEtAl2008}.
Here we take the above mentioned value adjusted to the experimental position of the Gamow-Teller
resonance in $^{208}$Pb within the relativistic (Q)RPA of Ref. \cite{PaarNiksicVretenarEtAl2004a}.

Matrix elements of the interaction of Eqs. (\ref{inter},\ref{LM}) in the particle-particle coupling scheme $\langle p_1n_2||{\tilde V}^{(pp)}||p_3n_4\rangle^{J}$ are related to the direct matrix elements used in the particle-hole coupling scheme $\langle p_1n_2||{\tilde V}^{(d)}||p_3n_4\rangle^{J}$ of Refs. \cite{NiksicMarketinVretenarEtAl2005,MarketinLitvinovaVretenarEtAl2012,RobinLitvinova2016} by the following recoupling relation:
\bea
\langle p_1n_2||{\tilde V}^{(pp)}||p_3n_4\rangle^{J} = \sum\limits_{\lambda} (2\lambda+1)(-1)^{j_3+j_4+J} \times \nonumber \\
\times \left\{\begin{array}{ccc}
j_1 & j_2 & J \\
j_3 & j_4 & \lambda \\
\end{array}\right\} \langle p_1n_2||{\tilde V}^{(d)}||p_3n_4\rangle^{\lambda}.             
\label{rec}
\eea
Thus, the pairing matrix elements are superpositions of the direct matrix elements with multipolarities allowed by the values of angular momenta of the proton-neutron pair configurations \cite{Serra2001}.  

The QHD for finite nuclei on the Hartree or Hartree-Fock level neglects time-dependence in the meson propagators of Eq. (\ref{inter}). The retardation effects of the in-medium nucleon-nucleon interaction can be taken into account rather accurately 
considering such temporal non-localities as coupling to small-amplitude vibrations of nucleonic density with various multipoles, as it can be shown consistently, for instance, in the equation of motion method \cite{AdachiSchuck1989,DukelskyRoepkeSchuck1998,SchuckTohyama2016a}.
The underlying mechanism of these vibrations is related to the meson exchange significantly modified by the nuclear medium. In the self-consistent theory these vibrations are correlated particle-hole or particle-particle (hole-hole) pairs generated by the same meson-exchange interaction $\tilde V$, and their characteristics can be calculated within the same linear response theory. These correlated two-fermion states provide a feedback on the single-particle degrees of freedom acting as mediators of an additional boson-exchange (commonly called phonon-exchange) interaction \cite{LitvinovaRing2006,Litvinova2012}.  Accounting for these contributions is a non-trivial task as it implies a non-perturbative solution of the BSE with a singular kernel. An approximation to such a solution based on the selection of the most relevant phonon-exchange contributions was proposed in Ref. \cite{Tselyaev1989a}, applied in \cite{KamerdzhievSpethTertychny2004,TselyaevSpethGrummerEtAl2007,LitvinovaTselyaev2007} and generalized to superfluid pairing in Refs. \cite{Tselyaev2007,LitvinovaTselyaev2007}, to the relativistic framework in Refs. \cite{LitvinovaRingTselyaev2007,LitvinovaRingTselyaev2008,LitvinovaRingTselyaev2010}, and to the spin-isospin response in Refs. \cite{MarketinLitvinovaVretenarEtAl2012,LitvinovaBrownFangEtAl2014,RobinLitvinova2016}. The main content of the approach is  the separation of $2n$-quasiparticle configurations according to their complexity based on the time projection technique in the many-body Green function formalism. According to its technique, the method was called time blocking approximation. Here we apply a similar technique to the proton-neutron particle-particle channel. In this approximation, confining by the pn$\otimes$phonon configurations without backward going terms, the dynamical (PVC) part of the interaction of Eq. (\ref{W-omega}) $\Phi(\omega)$ takes the form:
\begin{eqnarray}%
\Phi_{p_{1}n_{2},p_{3}n_{4}}^{\eta}(\omega)  = \nonumber \\
= \eta\sum\limits_{\mu}
\Bigl[\delta_{p_{1}p_{3}}\sum\limits_{n_{6}} \frac{\gamma_{\mu;n_{2}n_{6}%
}^{\eta} \gamma_{\mu;n_{4}n_{6}}^{\eta\ast}}{\omega-\varepsilon_{p_{1}}-\varepsilon_{n_{6}%
}-\eta\Omega_{\mu}} + \nonumber \\
+\ \delta_{n_{2}n_{4}}\sum\limits_{p_{5}}\frac{\gamma
_{\mu;p_{1}p_{5}}^{\eta}
\gamma_{\mu;p_{3}p_{5}}^{\eta\ast}}{\omega-
\varepsilon_{p_{5}} - \varepsilon_{n_{2}} - \eta\Omega_{\mu}} -\nonumber\\
-\Bigl(\frac{\gamma_{\mu;p_{1}p_{3}}^{\eta} \gamma_{\mu
;n_{4}n_{2}}^{\eta\ast}}{\omega- \varepsilon_{p_{3}}- \varepsilon_{n_{2}} -
\eta\Omega_{\mu}} +
\frac{\gamma_{\mu;p_{3}p_{1}}^{\eta\ast}\gamma_{\mu;n_{2}n_{4}}^{\eta}} {\omega- \varepsilon_{p_{1}} - \varepsilon_{n_{4}} -
\eta\Omega_{\mu}}\Bigr)\Bigr],\nonumber
\\
\label{phiphc0}%
\end{eqnarray} 
where we denote the matrix elements of the phonon-nucleon coupling vertices as $\gamma_{\mu;k_{1}k_{2}}^{\eta}$: 
\begin{equation}
\gamma_{\mu;k_{1}k_{2}}^{\eta} = \delta_{\eta,+1}\gamma_{\mu;k_{1}k_{2}} + \delta_{\eta,-1}\gamma_{\mu;k_{2}k_{1}}^{\ast}
\end{equation}
and their frequencies $\Omega_{\mu}$. The index $\mu$ numerates phonon quantum numbers, $\varepsilon_{k_{i}}$ are the energies of Dirac-Hartree  single-particle states, while indices $k_i$ stand for both proton and neutron levels. Here and in the following we will call the approach of Eq. (\ref{phiphc0}) to the interaction $\Phi(\omega)$ proton-neutron particle-particle relativistic time blocking approximation (pn-pp-RTBA), in analogy to the case of particle-hole pn-RTBA of Refs. \cite{MarketinLitvinovaVretenarEtAl2012,LitvinovaBrownFangEtAl2014}. 

The coupled form for the matrix elements of the amplitude $\Phi(\omega)$ for the pairing channel is obtained similarly to the particle-hole case \cite{LitvinovaRingTselyaev2007}. The phonon vertices $\gamma_{\mu;k_{2}k_{1}}$ are calculated within the relativistic random phase approximation (RRPA) \cite{RingMaVanGiaiEtAl2001}. In the present application we include only usual isoscalar phonons which are proven to play the leading role in the fragmentation of single-particle and collective states, although contributions from the isovector phonons can be also sizable \cite{Litvinova2016,RobinLitvinova2016}. 

The amplitude $\Phi(\omega)$ represents the energy-dependent part of the interaction and takes into account, as mentioned above, the retardation effects of the nucleon-nucleon interaction. In the lowest order time blocking approximation employed here, by making use of the time projection in the integral part of the BSE, pn$\otimes$phonon configurations are isolated and included consistently. Formally, the 
amplitude $\Phi(\omega)$ in Eq. (\ref{W-omega})  should be corrected by the subtraction of itself at zero frequency, in
order to avoid double counting of the phonon-nucleon coupling effects, which are implicitly included in the mean field \cite{Tselyaev2007,Tselyaev2013}. This was done in the previous studies of non-isospin-flip excitations. However, in the isospin-flip channels with unnatural parity transfer, where pion plays the major role, we do not perform the subtraction as pion does not contribute to the mean field on the Hartree level \cite{RobinLitvinova2016}. 
Generally, in the particle-hole channels  the amplitude $\Phi(\omega)$ causes an additional fragmentation and broadening of the strength functions \cite{LitvinovaRingTselyaev2007,LitvinovaRingTselyaev2008,MarketinLitvinovaVretenarEtAl2012,LitvinovaBrownFangEtAl2014,RobinLitvinova2016}. 
This is the consequence of the pole structure of this amplitude, in which the poles are located at the (Q)RPA particle-hole energy differences (or sums of the Bogoliubov quasiparticle energies in the case of pairing correlations) shifted by the phonon frequencies (see, for instance, Refs. \cite{LitvinovaRingTselyaev2007,RobinLitvinova2016}). In the present case the amplitude $\Phi(\omega)$ of Eq. (\ref{phiphc0}) has a similar form, but the particle-hole energy differences are replaced by the particle-particle and hole-hole energy sums, again the same as enter the uncorrelated proton-neutron (pn-RRPA) propagators of Eq. (\ref{R0}). Thus, a similar effect of fragmentation of the particle-particle and hole-hole states is expected from the qualitative point of view.   
Studies of the PVC effects on nuclear proton-neutron pairing properties are rather limited and so far mostly include those for the pairing gap in infinite nuclear matter \cite{ZhangCaoLombardoEtAl2016}. Therefore, here we are interested to see the quantitative PVC effects on the proton-neutron pairing strength and, in particular, to reveal if the PVC mechanism is favorable for the proton-neutron pairing.  

For this purpose, the microscopic strength function $S(E)$ is determined as a linear response to an external field $P$:
\begin{eqnarray}
S(E) &=& -\frac{1}{\pi}\lim\limits_{\Delta\rightarrow+0}Im\
\sum\limits_{p_{1}n_{2}p_{3}n_{4}}\sum\limits_{\eta\eta^{\prime}}
P_{p_{1}n_{2}}^{\eta\ast} \nonumber \\
&& \times R_{p_{1}n_{2},p_{3}n_{4}}^{\eta\eta^{\prime}}
(E + i\Delta)P_{p_{3}n_{4}}^{\eta^{\prime}},
\label{strf}%
\end{eqnarray}
where $P_{p_{1}n_{2}}^{\eta}$ is a matrix element of the spin-isospin-multipole operator
%
between proton-neutron particle for $\eta = 1$ (hole for $\eta = -1$) states corresponding to the proton-neutron addition (removal) fields, respectively.
In the calculations we will use a finite value of the imaginary part of the energy variable $\Delta$ = 200 keV for representation purposes.


\section{Details of calculations, results and discussion}
\label{sec:disc}
The BSE (\ref{respdir}) has been solved and the strength function (\ref{strf}) has been calculated for $J^{\pi} = 0^+$ and $J^{\pi} = 1^+$ proton-neutron pair transfers (removal and addition) for the doubly magic N=Z nuclei $^{56}$Ni and $^{100}$Sn. As mentioned above, in the present work we employ the NL3* parameter set for the relativistic mean-field, which provides the Dirac-Hartree single-nucleon basis states $\{p_i\},\{n_i\}$ for Eqs. (\ref{respdir}-\ref{strf}). While solving Eq. (\ref{respdir}), the basis in the Fermi sector is truncated in a way that the results at low energies are approximately saturated. It turns out that truncation at $|\varepsilon_{p_{i}}+\varepsilon_{n_{j}}| \leq$ 100 MeV for the particle-particle and hole-hole proton-neutron pairs entering Eq. (\ref{respdir}) is sufficient to provide stable results for the strength functions at low energies, except the cases of expected instabilities which will be discussed especially.  
The phonon space is truncated on  angular momenta, frequencies and reduced transition probabilities of the phonons and include isoscalar phonon modes with natural parities and multipolarities $J_{\mu} \leq$ 6, $\Omega_{\mu}\leq$ 20 MeV, and the reduced transition probabilities exceeding 5\% of the maximal one for the given multipolarity. This truncation scheme was used in the previous calculations for particle-hole types of nuclear response and found quite reasonable. For the $J^{\pi}=1^+$ channel PVC effects on the addition modes are included only up to 20 MeV of the excitation energy: at higher energies this is more demanding computationally. 

%
\begin{figure}
\begin{center}
\vspace{-0.5cm}
\includegraphics[scale=0.35]{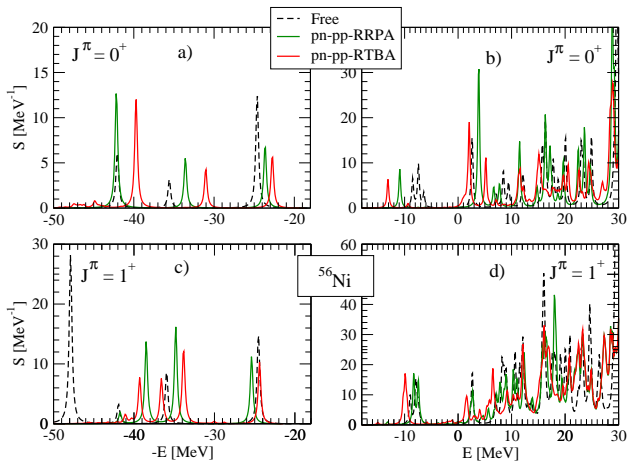}
\caption{Proton-neutron pair removal (a, c) and addition (b, d) strength distributions for $^{56}$Ni calculated with free response, proton-neutron particle-particle relativistic random phase approximation (pn-pp-RRPA) and in proton-neutron particle-particle relativistic time blocking approximation (pn-pp-RTBA).}
\label{56ni}
\end{center}
\end{figure}
\begin{figure}
\begin{center}
\vspace{-0.5cm}
\includegraphics[scale=0.35]{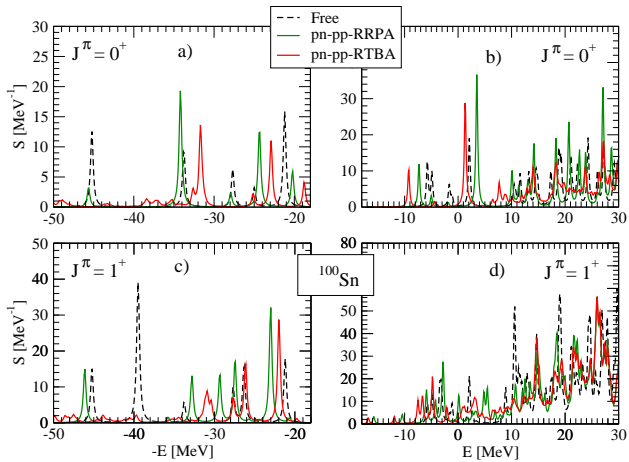}
\vspace{-1cm}
\caption{Same as in Fig. \ref{56ni} but for $^{100}$Sn.}
\label{100sn}%
\end{center}
\end{figure}
As follows from the definition of the pair transfer response, the resulting strength distributions represent the $J^{\pi}=0^+$ and $J^{\pi}=1^+$ spectra in (Z-1,N-1) and (Z+1,N+1) odd-odd nuclei.  
Figure \ref{56ni} shows those spectra with respect to the ground state (E = 0) of the mother nucleus $^{56}$Ni. The left panels a) and c) show the removal and the right panels b) and d) the addition modes for $J^{\pi} = 0^+$ (a, b) and $J^{\pi} = 1^+$ (c, d). The removal modes are plotted with the inverted sign of the energy axis. The removal modes should be understood as the ground and excited states of $^{54}$Co  while the addition modes as those of $^{58}$Cu.  The spectra of Fig. \ref{56ni} have been calculated in the following three approximations: (i) "free" response neglecting all the residual interaction effects (black dashed curves), (ii) proton-neutron particle-particle RRPA (pn-pp-RRPA), which includes the effects of the static interaction ${\tilde V}$, but neglects the retardation effects of the second term of Eq. (\ref{W-omega}) (green curves), and (iii) proton-neutron particle-particle RTBA (pn-pp-RTBA) with the full interaction of Eq. (\ref{W-omega}) (red curves). Comparison between the free response and the pn-pp-RRPA one allows an assessment of the effect of the meson exchange interaction including the parity-conserving $\rho$-meson exchange and parity-breaking pion exchange in the leading approximation. 

The general picture of the resulting removal and addition spectra admits identification of the ground-state (the lowest) solutions, soft low-lying modes and high-frequency oscillations, which are often called giant pairing vibrations, the analogs of  giant resonances of the particle-hole type.  In the following discussion we will mostly focus on the ground-state and low-energy solutions. Comparing the free response and pn-pp-RRPA strengths for the $J^{\pi}=0^+$ removal and addition modes shown in panels (a) and (b) of Fig. \ref{56ni}, one can observe relatively strong shifts downward of the lowest solutions of the BSE. Some of the higher-energy solutions, however, move upward with an enhanced strength. This indicates an interplay of attractive and repulsive contributions in the static part of the pairing interaction and its potential to form collective states in the proton-neutron pairing channel. The next observation comes from comparing the strength functions 
calculated in pn-pp-RRPA (green) and pn-pp-RTBA (red), in order to see the effect of PVC on these modes of excitation. As expected, one observes a typical PVC-induced fragmentation of the spectra and shifting the lowest solutions further down in energy in the latter case. The effects of both the meson-exchange and the PVC on the lowest states are stronger for the addition modes: each of them introduces $\approx$ 2 MeV shifts while the corresponding shifts in the removal sector are twice smaller.  

The $J^{\pi}=1^+$ response functions displayed in panels (c) and (d) of Fig. \ref{56ni} show somewhat different trends. In the pair removal sector, the meson-exchange interaction acts repulsively on the lowest state while PVC brings it back down 100 keV below its unperturbed location. At higher energies one can observe attractive action of the meson exchange and some fragmentation effect of PVC. In the addition sector, the pn-pp-RRPA splits out a relatively weak state which is shifted by 4 MeV down from the lowest unperturbed level. PVC acts in the same direction and causes further shifting down of the lowest state as well as the next low-lying mode which comes out 10 MeV below the ground state of $^{56}$Ni. States at higher energies are overall fragmented due to the PVC mechanism. Here in $J^{\pi}=1^+$ pair addition sector  we have observed an instability of the lowest solution, for instance, relatively modest changes in the truncation scheme, such as the particle-particle or phonon basis extension cause further movement down of this solution. This indicates the tendency to form the deuteron condensate phase. 

The ground state of the nucleus $^{58}$Cu is known to have spin and parity $1^+$ and located by 13.11 MeV lower than the ground state of $^{56}$Ni, and the first $0^+$ state is just 203 keV above it \cite{AudiWapstraThibault2003,a}. As follows from Fig. \ref{56ni}, pn-pp-RRPA gives $J^{\pi} = 1^+$ and the location of the lowest state in $^{58}$Cu at -13.2 MeV while in pn-pp-RTBA this state is shifted down by 2 MeV, with respect to the ground state of $^{56}$Ni. However, because of the above mentioned instability, these values have large uncertainties. To avoid such instabilities, the approach should be modified as discussed, for instance, in Ref. \cite{Dickhoff2005}. This will be done elsewhere. 

In the hole-hole branch of the removal modes the lowest-energy states correspond to the ground and first excited states  of (N-1,Z-1) nuclei. Thus, starting from $^{56}$Ni core, we arrive to $^{54}$Co whose experimental ground state has $J^{\pi} = 0^+$ and the location 21.26 MeV above the ground state of $^{56}$Ni \cite{AudiWapstraThibault2003}. The lowest $1^+$ state in $^{54}$Co is 937 keV above its ground state \cite{a}. Our calculations also point out that the ground state of $^{54}$Co has spin and parity $0^+$, and the energy 23.6 MeV in pn-pp-RTBA and 22.7 MeV in pn-pp-RTBA with respect to the ground state of $^{56}$Ni, i.e. the PVC brings the ground state energy of $^{54}$Co to a better agreement with data. The first $J^{\pi}=1^{+}$ state is 1.7 MeV above the ground state according to pn-pp-RTBA, which is a rather reasonable value.

A similar situation is found for $^{100}$Sn, as displayed in Figure \ref{100sn}. For the $0^+$ modes both meson-exchange and PVC interactions produce downward shifts of the lowest state with respect to its unperturbed  location, in both removal and addition channels. The $1^+$ spectra show a similar trend, however, the lowest $1^+$ solutions are rather weak and,  like in the case of $^{56}$Ni, somewhat unstable in the addition sector. Again, PVC further lowers the lowest solutions and generates an overall fragmentation of the higher-lying states.
Experimental information about odd-odd nuclei around $^{100}$Sn is very limited. The nucleus $^{102}$Sb is experimentally unknown, so that it is not possible to assess the performance of our approach for the addition modes. From Ref. \cite{AudiWapstraThibault2003} one can extract the binding energy of $^{98}$In which indicates that its ground state is 18.26 MeV above the one of $^{100}$Sn. The spin of $^{98}$In ground state is, however, undefined \cite{a}. In our pn-pp-RTBA calculations shown in  Fig. \ref{100sn} the lowest $0^{+}$ state of $^{98}$In comes out below $1^+_1$ and 18.8 MeV above the ground state of $^{100}$Sn. 
Thus, it can be a candidate for being the ground state of $^{98}$In. 
To have a more complete picture of the spectra, we would need to perform calculations for other spins and parities $J^{\pi}$, which may be done elsewhere.

\section{Summary}
\label{summary}
In this work, we have formulated proton-neutron particle-particle relativistic time blocking approximation and investigated its potential of describing low-lying $J^{\pi} = 0^+$ and $J^{\pi} =1^+$ states in N=Z odd-odd medium-mass nuclei near the shell closures. For this purpose,  
we have computed and analyzed proton-neutron pair addition and removal modes of excitation in doubly-magic medium-mass nuclei $^{56}$Ni and $^{100}$Sn. We have found that quite a delicate balance between attractive and repulsive mechanisms in the meson-exchange sector, together with the core polarization effects of the particle-vibration coupling, may lead to a triplet deuteron condensate formation in medium-mass N=Z odd-odd nuclei around $^{56}$Ni and $^{100}$Sn. 

Another conclusion from this work is that core polarization and retardation effects associated with the particle-vibration coupling play an important role in the proton-neutron pairing channel. PVC brings the location of the lowest-energy states of odd-odd nuclei to a better agreement with data, as compared to calculations which include only the static part of the nucleon-nucleon interaction, and may reinforce the deuteron condensate formation.

More generally, the soft modes in the proton-neutron pairing channel are shown to be of interest as precursory phenomena and as potential mediators of dynamical proton-neutron pairing. 
Therefore, further studies can include a more systematic investigation of N=Z nuclei including also like-particle pairing, calculations of higher multipoles in the proton-neutron pairing channel and studies of their potential influence on other isospin-flip excitations, such as Gamow-Teller resonance, spin-dipole resonance and others, which are known to be sensitive to proton-neutron pairing.

%
\section*{Acknowledgements}
%
%
The authors greatly appreciate discussions with A.V. Afanasjev, U. Lombardo, P. Ring and P. Schuck. Special thanks to T. Marketin for providing a part of the code for pn-RRPA matrix elements. This work is partly supported by US-NSF grant PHY-1404343 and by NSF Career grant PHY-1654379. Support by the Institute for Nuclear Theory under US-DOE Grant DE-FG02-00ER41132 and by JINA-CEE under US-NSF Grant PHY-1430152 is also acknowledged.


\bibliography{Bibliography_Nov2016}

\begin{thebibliography}{64}%
\makeatletter
\providecommand \@ifxundefined [1]{%
 \@ifx{#1\undefined}
}%
\providecommand \@ifnum [1]{%
 \ifnum #1\expandafter \@firstoftwo
 \else \expandafter \@secondoftwo
 \fi
}%
\providecommand \@ifx [1]{%
 \ifx #1\expandafter \@firstoftwo
 \else \expandafter \@secondoftwo
 \fi
}%
\providecommand \natexlab [1]{#1}%
\providecommand \enquote  [1]{``#1''}%
\providecommand \bibnamefont  [1]{#1}%
\providecommand \bibfnamefont [1]{#1}%
\providecommand \citenamefont [1]{#1}%
\providecommand \href@noop [0]{\@secondoftwo}%
\providecommand \href [0]{\begingroup \@sanitize@url \@href}%
\providecommand \@href[1]{\@@startlink{#1}\@@href}%
\providecommand \@@href[1]{\endgroup#1\@@endlink}%
\providecommand \@sanitize@url [0]{\catcode `\\12\catcode `\$12\catcode
  `\&12\catcode `\#12\catcode `\^12\catcode `\_12\catcode `\%12\relax}%
\providecommand \@@startlink[1]{}%
\providecommand \@@endlink[0]{}%
\providecommand \url  [0]{\begingroup\@sanitize@url \@url }%
\providecommand \@url [1]{\endgroup\@href {#1}{\urlprefix }}%
\providecommand \urlprefix  [0]{URL }%
\providecommand \Eprint [0]{\href }%
\providecommand \doibase [0]{http://dx.doi.org/}%
\providecommand \selectlanguage [0]{\@gobble}%
\providecommand \bibinfo  [0]{\@secondoftwo}%
\providecommand \bibfield  [0]{\@secondoftwo}%
\providecommand \translation [1]{[#1]}%
\providecommand \BibitemOpen [0]{}%
\providecommand \bibitemStop [0]{}%
\providecommand \bibitemNoStop [0]{.\EOS\space}%
\providecommand \EOS [0]{\spacefactor3000\relax}%
\providecommand \BibitemShut  [1]{\csname bibitem#1\endcsname}%
\let\auto@bib@innerbib\@empty
\bibitem [{\citenamefont {Ring}\ and\ \citenamefont
  {Schuck}(1980)}]{RingSchuck1980}%
  \BibitemOpen
  \bibfield  {author} {\bibinfo {author} {\bibfnamefont {P.}~\bibnamefont
  {Ring}}\ and\ \bibinfo {author} {\bibfnamefont {P.}~\bibnamefont {Schuck}},\
  }\href@noop {} {\emph {\bibinfo {title} {The Nuclear Many-Body Problem}}}\
  (\bibinfo  {publisher} {Springer-Verlag Berlin Heidelberg},\ \bibinfo {year}
  {1980})\BibitemShut {NoStop}%
\bibitem [{\citenamefont {Bohr}\ \emph {et~al.}(1958)\citenamefont {Bohr},
  \citenamefont {Mottelson},\ and\ \citenamefont
  {Pines}}]{BohrMottelsonPines1958}%
  \BibitemOpen
  \bibfield  {author} {\bibinfo {author} {\bibfnamefont {A.}~\bibnamefont
  {Bohr}}, \bibinfo {author} {\bibfnamefont {B.}~\bibnamefont {Mottelson}}, \
  and\ \bibinfo {author} {\bibfnamefont {D.}~\bibnamefont {Pines}},\
  }\href@noop {} {\bibfield  {journal} {\bibinfo  {journal} {Physical Review}\
  }\textbf {\bibinfo {volume} {110}},\ \bibinfo {pages} {936} (\bibinfo {year}
  {1958})}\BibitemShut {NoStop}%
\bibitem [{\citenamefont {Broglia}\ and\ \citenamefont
  {Zelevinsky}(2013)}]{50BCS}%
  \BibitemOpen
  \bibinfo {editor} {\bibfnamefont {R.}~\bibnamefont {Broglia}}\ and\ \bibinfo
  {editor} {\bibfnamefont {V.}~\bibnamefont {Zelevinsky}},\ eds.,\ \href@noop
  {} {\emph {\bibinfo {title} {Fifty Years Of Nuclear BCS: Pairing In Finite
  Systems}}}\ (\bibinfo  {publisher} {World Scientific},\ \bibinfo {year}
  {2013})\BibitemShut {NoStop}%
\bibitem [{\citenamefont {Barranco}\ \emph {et~al.}(1999)\citenamefont
  {Barranco}, \citenamefont {Broglia}, \citenamefont {Gori}, \citenamefont
  {Vigezzi}, \citenamefont {Bortignon},\ and\ \citenamefont
  {Terasaki}}]{BarrancoBrogliaGoriEtAl1999}%
  \BibitemOpen
  \bibfield  {author} {\bibinfo {author} {\bibfnamefont {F.}~\bibnamefont
  {Barranco}}, \bibinfo {author} {\bibfnamefont {R.}~\bibnamefont {Broglia}},
  \bibinfo {author} {\bibfnamefont {G.}~\bibnamefont {Gori}}, \bibinfo {author}
  {\bibfnamefont {E.}~\bibnamefont {Vigezzi}}, \bibinfo {author} {\bibfnamefont
  {P.}~\bibnamefont {Bortignon}}, \ and\ \bibinfo {author} {\bibfnamefont
  {J.}~\bibnamefont {Terasaki}},\ }\href@noop {} {\bibfield  {journal}
  {\bibinfo  {journal} {Physical Review Letters}\ }\textbf {\bibinfo {volume}
  {83}},\ \bibinfo {pages} {2147} (\bibinfo {year} {1999})}\BibitemShut
  {NoStop}%
\bibitem [{\citenamefont {Avdeenkov}\ and\ \citenamefont
  {Kamerdzhiev}(1999)}]{AvdeenkovKamerdzhiev1999}%
  \BibitemOpen
  \bibfield  {author} {\bibinfo {author} {\bibfnamefont {A.~V.}\ \bibnamefont
  {Avdeenkov}}\ and\ \bibinfo {author} {\bibfnamefont {S.~P.}\ \bibnamefont
  {Kamerdzhiev}},\ }\href@noop {} {\bibfield  {journal} {\bibinfo  {journal}
  {Physics of Atomic Nuclei}\ }\textbf {\bibinfo {volume} {62}},\ \bibinfo
  {pages} {563} (\bibinfo {year} {1999})}\BibitemShut {NoStop}%
\bibitem [{\citenamefont {Barranco}\ \emph {et~al.}(2005)\citenamefont
  {Barranco}, \citenamefont {Bortignon}, \citenamefont {Broglia}, \citenamefont
  {Col{\`o}}, \citenamefont {Schuck}, \citenamefont {Vigezzi},\ and\
  \citenamefont {Vinas}}]{BarrancoBortignonBrogliaEtAl2005}%
  \BibitemOpen
  \bibfield  {author} {\bibinfo {author} {\bibfnamefont {F.}~\bibnamefont
  {Barranco}}, \bibinfo {author} {\bibfnamefont {P.}~\bibnamefont {Bortignon}},
  \bibinfo {author} {\bibfnamefont {R.}~\bibnamefont {Broglia}}, \bibinfo
  {author} {\bibfnamefont {G.}~\bibnamefont {Col{\`o}}}, \bibinfo {author}
  {\bibfnamefont {P.}~\bibnamefont {Schuck}}, \bibinfo {author} {\bibfnamefont
  {E.}~\bibnamefont {Vigezzi}}, \ and\ \bibinfo {author} {\bibfnamefont
  {X.}~\bibnamefont {Vinas}},\ }\href@noop {} {\bibfield  {journal} {\bibinfo
  {journal} {Phys. Rev. C}\ }\textbf {\bibinfo {volume} {72}},\ \bibinfo
  {pages} {054314} (\bibinfo {year} {2005})}\BibitemShut {NoStop}%
\bibitem [{\citenamefont {Idini}\ \emph {et~al.}(2015)\citenamefont {Idini},
  \citenamefont {Potel}, \citenamefont {Barranco}, \citenamefont {Vigezzi},\
  and\ \citenamefont {Broglia}}]{IdiniPotelBarrancoEtAl2015}%
  \BibitemOpen
  \bibfield  {author} {\bibinfo {author} {\bibfnamefont {A.}~\bibnamefont
  {Idini}}, \bibinfo {author} {\bibfnamefont {G.}~\bibnamefont {Potel}},
  \bibinfo {author} {\bibfnamefont {F.}~\bibnamefont {Barranco}}, \bibinfo
  {author} {\bibfnamefont {E.}~\bibnamefont {Vigezzi}}, \ and\ \bibinfo
  {author} {\bibfnamefont {R.~A.}\ \bibnamefont {Broglia}},\ }\href@noop {}
  {\bibfield  {journal} {\bibinfo  {journal} {Physical Review C}\ }\textbf
  {\bibinfo {volume} {92}},\ \bibinfo {pages} {031304} (\bibinfo {year}
  {2015})}\BibitemShut {NoStop}%
\bibitem [{\citenamefont {B\`es}\ and\ \citenamefont
  {Broglia}(1966)}]{BesBroglia1966}%
  \BibitemOpen
  \bibfield  {author} {\bibinfo {author} {\bibfnamefont {D.~R.}\ \bibnamefont
  {B\`es}}\ and\ \bibinfo {author} {\bibfnamefont {R.~A.}\ \bibnamefont
  {Broglia}},\ }\href@noop {} {\bibfield  {journal} {\bibinfo  {journal}
  {Nuclear Physics}\ }\textbf {\bibinfo {volume} {80}},\ \bibinfo {pages} {289}
  (\bibinfo {year} {1966})}\BibitemShut {NoStop}%
\bibitem [{\citenamefont {Ripka}\ and\ \citenamefont
  {Padjen}(1969)}]{RipkaPadjen1969}%
  \BibitemOpen
  \bibfield  {author} {\bibinfo {author} {\bibfnamefont {G.}~\bibnamefont
  {Ripka}}\ and\ \bibinfo {author} {\bibfnamefont {R.}~\bibnamefont {Padjen}},\
  }\href@noop {} {\bibfield  {journal} {\bibinfo  {journal} {Nuclear Physics
  A}\ }\textbf {\bibinfo {volume} {132}},\ \bibinfo {pages} {489} (\bibinfo
  {year} {1969})}\BibitemShut {NoStop}%
\bibitem [{\citenamefont {Hahne}\ \emph {et~al.}(1977)\citenamefont {Hahne},
  \citenamefont {Heiss},\ and\ \citenamefont
  {Engelbrecht}}]{HahneHeissEngelbrecht1977}%
  \BibitemOpen
  \bibfield  {author} {\bibinfo {author} {\bibfnamefont {F.~J.~W.}\
  \bibnamefont {Hahne}}, \bibinfo {author} {\bibfnamefont {W.~D.}\ \bibnamefont
  {Heiss}}, \ and\ \bibinfo {author} {\bibfnamefont {C.~A.}\ \bibnamefont
  {Engelbrecht}},\ }\href {\doibase 10.1016/0003-4916(77)90332-3} {\bibfield
  {journal} {\bibinfo  {journal} {Annals of Physics}\ }\textbf {\bibinfo
  {volume} {104}},\ \bibinfo {pages} {251} (\bibinfo {year}
  {1977})}\BibitemShut {NoStop}%
\bibitem [{\citenamefont {Engel}\ \emph {et~al.}(1996)\citenamefont {Engel},
  \citenamefont {Langanke},\ and\ \citenamefont
  {Vogel}}]{EngelLangankeVogel1996}%
  \BibitemOpen
  \bibfield  {author} {\bibinfo {author} {\bibfnamefont {J.}~\bibnamefont
  {Engel}}, \bibinfo {author} {\bibfnamefont {K.}~\bibnamefont {Langanke}}, \
  and\ \bibinfo {author} {\bibfnamefont {P.}~\bibnamefont {Vogel}},\ }\href
  {\doibase 10.1016/S0370-2693(96)01294-4} {\bibfield  {journal} {\bibinfo
  {journal} {Physics Letters B}\ }\textbf {\bibinfo {volume} {389}},\ \bibinfo
  {pages} {211} (\bibinfo {year} {1996})}\BibitemShut {NoStop}%
\bibitem [{\citenamefont {Langanke}\ and\ \citenamefont
  {Mart\'inez-Pinedo}(2013)}]{LangankeMartinez-Pinedo2013}%
  \BibitemOpen
  \bibfield  {author} {\bibinfo {author} {\bibfnamefont {K.}~\bibnamefont
  {Langanke}}\ and\ \bibinfo {author} {\bibfnamefont {G.}~\bibnamefont
  {Mart\'inez-Pinedo}},\ }\href@noop {} {\emph {\bibinfo {title} {Fifty Years
  of Nuclear BCS, Eds. R. A. Broglia and V. G. Zelevinsky}}}\ (\bibinfo
  {publisher} {World Scientific},\ \bibinfo {year} {2013})\ Chap.~\bibinfo
  {chapter} {12}, p.\ \bibinfo {pages} {154}\BibitemShut {NoStop}%
\bibitem [{\citenamefont {Satula}\ and\ \citenamefont
  {Wyss}(1997)}]{SatulaWyss1997}%
  \BibitemOpen
  \bibfield  {author} {\bibinfo {author} {\bibfnamefont {W.}~\bibnamefont
  {Satula}}\ and\ \bibinfo {author} {\bibfnamefont {R.}~\bibnamefont {Wyss}},\
  }\href@noop {} {\bibfield  {journal} {\bibinfo  {journal} {Physics Letters}\
  }\textbf {\bibinfo {volume} {B393}},\ \bibinfo {pages} {1} (\bibinfo {year}
  {1997})}\BibitemShut {NoStop}%
\bibitem [{\citenamefont {Goodman}(1999)}]{Goodman1999}%
  \BibitemOpen
  \bibfield  {author} {\bibinfo {author} {\bibfnamefont {A.~L.}\ \bibnamefont
  {Goodman}},\ }\href@noop {} {\bibfield  {journal} {\bibinfo  {journal}
  {Physical Review C}\ }\textbf {\bibinfo {volume} {60}},\ \bibinfo {pages}
  {014311} (\bibinfo {year} {1999})}\BibitemShut {NoStop}%
\bibitem [{\citenamefont {Bertsch}\ and\ \citenamefont
  {Luo}(2010)}]{BertschLuo2010}%
  \BibitemOpen
  \bibfield  {author} {\bibinfo {author} {\bibfnamefont {G.~F.}\ \bibnamefont
  {Bertsch}}\ and\ \bibinfo {author} {\bibfnamefont {Y.}~\bibnamefont {Luo}},\
  }\href@noop {} {\bibfield  {journal} {\bibinfo  {journal} {Physical Review
  C}\ }\textbf {\bibinfo {volume} {81}},\ \bibinfo {pages} {064320} (\bibinfo
  {year} {2010})}\BibitemShut {NoStop}%
\bibitem [{\citenamefont {Gezerlis}\ \emph {et~al.}(2011)\citenamefont
  {Gezerlis}, \citenamefont {Bertsch},\ and\ \citenamefont
  {Luo}}]{GezerlisBertschLuo2011a}%
  \BibitemOpen
  \bibfield  {author} {\bibinfo {author} {\bibfnamefont {A.}~\bibnamefont
  {Gezerlis}}, \bibinfo {author} {\bibfnamefont {G.~F.}\ \bibnamefont
  {Bertsch}}, \ and\ \bibinfo {author} {\bibfnamefont {Y.~L.}\ \bibnamefont
  {Luo}},\ }\href@noop {} {\bibfield  {journal} {\bibinfo  {journal} {Physical
  Review Letters}\ }\textbf {\bibinfo {volume} {106}},\ \bibinfo {pages}
  {252502} (\bibinfo {year} {2011})}\BibitemShut {NoStop}%
\bibitem [{\citenamefont {Yoshida}(2014)}]{Yoshida2014}%
  \BibitemOpen
  \bibfield  {author} {\bibinfo {author} {\bibfnamefont {K.}~\bibnamefont
  {Yoshida}},\ }\href@noop {} {\bibfield  {journal} {\bibinfo  {journal}
  {Physical Review}\ }\textbf {\bibinfo {volume} {C90}},\ \bibinfo {pages}
  {031303} (\bibinfo {year} {2014})}\BibitemShut {NoStop}%
\bibitem [{\citenamefont {Frauendorf}\ and\ \citenamefont
  {Macchiavelli}(2014)}]{Frauendorf201424}%
  \BibitemOpen
  \bibfield  {author} {\bibinfo {author} {\bibfnamefont {S.}~\bibnamefont
  {Frauendorf}}\ and\ \bibinfo {author} {\bibfnamefont {A.}~\bibnamefont
  {Macchiavelli}},\ }\href {\doibase
  http://dx.doi.org/10.1016/j.ppnp.2014.07.001} {\bibfield  {journal} {\bibinfo
   {journal} {Progress in Particle and Nuclear Physics}\ }\textbf {\bibinfo
  {volume} {78}},\ \bibinfo {pages} {24 } (\bibinfo {year} {2014})}\BibitemShut
  {NoStop}%
\bibitem [{\citenamefont {Zhang}\ \emph {et~al.}(2016)\citenamefont {Zhang},
  \citenamefont {Cao}, \citenamefont {Lombardo},\ and\ \citenamefont
  {Schuck}}]{ZhangCaoLombardoEtAl2016}%
  \BibitemOpen
  \bibfield  {author} {\bibinfo {author} {\bibfnamefont {S.~S.}\ \bibnamefont
  {Zhang}}, \bibinfo {author} {\bibfnamefont {L.~G.}\ \bibnamefont {Cao}},
  \bibinfo {author} {\bibfnamefont {U.}~\bibnamefont {Lombardo}}, \ and\
  \bibinfo {author} {\bibfnamefont {P.}~\bibnamefont {Schuck}},\ }\href@noop {}
  {\bibfield  {journal} {\bibinfo  {journal} {Physical Review C}\ }\textbf
  {\bibinfo {volume} {93}},\ \bibinfo {pages} {044329} (\bibinfo {year}
  {2016})}\BibitemShut {NoStop}%
\bibitem [{\citenamefont {Robin}\ and\ \citenamefont
  {Litvinova}(2016)}]{RobinLitvinova2016}%
  \BibitemOpen
  \bibfield  {author} {\bibinfo {author} {\bibfnamefont {C.}~\bibnamefont
  {Robin}}\ and\ \bibinfo {author} {\bibfnamefont {E.}~\bibnamefont
  {Litvinova}},\ }\href@noop {} {\bibfield  {journal} {\bibinfo  {journal}
  {European Physical Journal A}\ }\textbf {\bibinfo {volume} {52}},\ \bibinfo
  {pages} {205} (\bibinfo {year} {2016})}\BibitemShut {NoStop}%
\bibitem [{\citenamefont {Nik\ifmmode \check{s}\else
  \v{s}\fi{}i\ifmmode~\acute{c}\else \'{c}\fi{}}\ \emph
  {et~al.}(2005)\citenamefont {Nik\ifmmode \check{s}\else
  \v{s}\fi{}i\ifmmode~\acute{c}\else \'{c}\fi{}}, \citenamefont {Marketin},
  \citenamefont {Vretenar}, \citenamefont {Paar},\ and\ \citenamefont
  {Ring}}]{NiksicMarketinVretenarEtAl2005}%
  \BibitemOpen
  \bibfield  {author} {\bibinfo {author} {\bibfnamefont {T.}~\bibnamefont
  {Nik\ifmmode \check{s}\else \v{s}\fi{}i\ifmmode~\acute{c}\else \'{c}\fi{}}},
  \bibinfo {author} {\bibfnamefont {T.}~\bibnamefont {Marketin}}, \bibinfo
  {author} {\bibfnamefont {D.}~\bibnamefont {Vretenar}}, \bibinfo {author}
  {\bibfnamefont {N.}~\bibnamefont {Paar}}, \ and\ \bibinfo {author}
  {\bibfnamefont {P.}~\bibnamefont {Ring}},\ }\href@noop {} {\bibfield
  {journal} {\bibinfo  {journal} {Physical Review C}\ }\textbf {\bibinfo
  {volume} {71}},\ \bibinfo {pages} {014308} (\bibinfo {year}
  {2005})}\BibitemShut {NoStop}%
\bibitem [{\citenamefont {Engel}\ \emph {et~al.}(1999)\citenamefont {Engel},
  \citenamefont {Bender}, \citenamefont {Dobaczewski}, \citenamefont
  {Nazarewicz},\ and\ \citenamefont {Surman}}]{EngelBenderDobaczewskiEtAl1999}%
  \BibitemOpen
  \bibfield  {author} {\bibinfo {author} {\bibfnamefont {J.}~\bibnamefont
  {Engel}}, \bibinfo {author} {\bibfnamefont {M.}~\bibnamefont {Bender}},
  \bibinfo {author} {\bibfnamefont {J.}~\bibnamefont {Dobaczewski}}, \bibinfo
  {author} {\bibfnamefont {W.}~\bibnamefont {Nazarewicz}}, \ and\ \bibinfo
  {author} {\bibfnamefont {R.}~\bibnamefont {Surman}},\ }\href@noop {}
  {\bibfield  {journal} {\bibinfo  {journal} {Physical Review C}\ }\textbf
  {\bibinfo {volume} {60}},\ \bibinfo {pages} {014302} (\bibinfo {year}
  {1999})}\BibitemShut {NoStop}%
\bibitem [{\citenamefont {Litvinova}(2016)}]{Litvinova2016}%
  \BibitemOpen
  \bibfield  {author} {\bibinfo {author} {\bibfnamefont {E.}~\bibnamefont
  {Litvinova}},\ }\href {\doibase
  http://dx.doi.org/10.1016/j.physletb.2016.01.052} {\bibfield  {journal}
  {\bibinfo  {journal} {Physics Letters B}\ }\textbf {\bibinfo {volume}
  {755}},\ \bibinfo {pages} {138 } (\bibinfo {year} {2016})}\BibitemShut
  {NoStop}%
\bibitem [{\citenamefont {Robin}\ and\ \citenamefont
  {Litvinova}(2017)}]{RobinLitvinova2017}%
  \BibitemOpen
  \bibfield  {author} {\bibinfo {author} {\bibfnamefont {C.}~\bibnamefont
  {Robin}}\ and\ \bibinfo {author} {\bibfnamefont {E.}~\bibnamefont
  {Litvinova}},\ }\href@noop {} {\ \textbf {\bibinfo {volume} {PoS INPC2016}},\
  \bibinfo {pages} {020} (\bibinfo {year} {2017})}\BibitemShut {NoStop}%
\bibitem [{\citenamefont {Litvinova}\ \emph {et~al.}(2007)\citenamefont
  {Litvinova}, \citenamefont {Ring},\ and\ \citenamefont
  {Tselyaev}}]{LitvinovaRingTselyaev2007}%
  \BibitemOpen
  \bibfield  {author} {\bibinfo {author} {\bibfnamefont {E.}~\bibnamefont
  {Litvinova}}, \bibinfo {author} {\bibfnamefont {P.}~\bibnamefont {Ring}}, \
  and\ \bibinfo {author} {\bibfnamefont {V.}~\bibnamefont {Tselyaev}},\
  }\href@noop {} {\bibfield  {journal} {\bibinfo  {journal} {Physical Review
  C}\ }\textbf {\bibinfo {volume} {75}},\ \bibinfo {pages} {064308} (\bibinfo
  {year} {2007})}\BibitemShut {NoStop}%
\bibitem [{\citenamefont {Marketin}\ \emph {et~al.}(2012)\citenamefont
  {Marketin}, \citenamefont {Litvinova}, \citenamefont {Vretenar},\ and\
  \citenamefont {Ring}}]{MarketinLitvinovaVretenarEtAl2012}%
  \BibitemOpen
  \bibfield  {author} {\bibinfo {author} {\bibfnamefont {T.}~\bibnamefont
  {Marketin}}, \bibinfo {author} {\bibfnamefont {E.}~\bibnamefont {Litvinova}},
  \bibinfo {author} {\bibfnamefont {D.}~\bibnamefont {Vretenar}}, \ and\
  \bibinfo {author} {\bibfnamefont {P.}~\bibnamefont {Ring}},\ }\href@noop {}
  {\bibfield  {journal} {\bibinfo  {journal} {Physics Letters B}\ }\textbf
  {\bibinfo {volume} {706}},\ \bibinfo {pages} {477} (\bibinfo {year}
  {2012})}\BibitemShut {NoStop}%
\bibitem [{\citenamefont {Litvinova}\ \emph {et~al.}(2014)\citenamefont
  {Litvinova}, \citenamefont {Brown}, \citenamefont {Fang}, \citenamefont
  {Marketin},\ and\ \citenamefont {Zegers}}]{LitvinovaBrownFangEtAl2014}%
  \BibitemOpen
  \bibfield  {author} {\bibinfo {author} {\bibfnamefont {E.}~\bibnamefont
  {Litvinova}}, \bibinfo {author} {\bibfnamefont {B.}~\bibnamefont {Brown}},
  \bibinfo {author} {\bibfnamefont {D.-L.}\ \bibnamefont {Fang}}, \bibinfo
  {author} {\bibfnamefont {T.}~\bibnamefont {Marketin}}, \ and\ \bibinfo
  {author} {\bibfnamefont {R.}~\bibnamefont {Zegers}},\ }\href@noop {}
  {\bibfield  {journal} {\bibinfo  {journal} {Physics Letters B}\ }\textbf
  {\bibinfo {volume} {730}},\ \bibinfo {pages} {307} (\bibinfo {year}
  {2014})}\BibitemShut {NoStop}%
\bibitem [{\citenamefont {Ring}(1996)}]{Ring1996}%
  \BibitemOpen
  \bibfield  {author} {\bibinfo {author} {\bibfnamefont {P.}~\bibnamefont
  {Ring}},\ }\href@noop {} {\bibfield  {journal} {\bibinfo  {journal} {Progress
  in Particle and Nuclear Physics}\ }\textbf {\bibinfo {volume} {37}},\
  \bibinfo {pages} {193} (\bibinfo {year} {1996})}\BibitemShut {NoStop}%
\bibitem [{\citenamefont {Vretenar}\ \emph {et~al.}(2005)\citenamefont
  {Vretenar}, \citenamefont {Afanasjev}, \citenamefont {Lalazissis},\ and\
  \citenamefont {Ring}}]{VretenarAfanasjevLalazissisEtAl2005}%
  \BibitemOpen
  \bibfield  {author} {\bibinfo {author} {\bibfnamefont {D.}~\bibnamefont
  {Vretenar}}, \bibinfo {author} {\bibfnamefont {A.~V.}\ \bibnamefont
  {Afanasjev}}, \bibinfo {author} {\bibfnamefont {G.~A.}\ \bibnamefont
  {Lalazissis}}, \ and\ \bibinfo {author} {\bibfnamefont {P.}~\bibnamefont
  {Ring}},\ }\href@noop {} {\bibfield  {journal} {\bibinfo  {journal} {Physics
  Reports}\ }\textbf {\bibinfo {volume} {409}},\ \bibinfo {pages} {101}
  (\bibinfo {year} {2005})}\BibitemShut {NoStop}%
\bibitem [{\citenamefont {Liang}\ \emph {et~al.}(2008)\citenamefont {Liang},
  \citenamefont {Van~Giai},\ and\ \citenamefont
  {Meng}}]{LiangVanGiaiMengEtAl2008}%
  \BibitemOpen
  \bibfield  {author} {\bibinfo {author} {\bibfnamefont {H.}~\bibnamefont
  {Liang}}, \bibinfo {author} {\bibfnamefont {N.}~\bibnamefont {Van~Giai}}, \
  and\ \bibinfo {author} {\bibfnamefont {J.}~\bibnamefont {Meng}},\ }\href@noop
  {} {\bibfield  {journal} {\bibinfo  {journal} {Physical Review Letters}\
  }\textbf {\bibinfo {volume} {101}},\ \bibinfo {pages} {122502} (\bibinfo
  {year} {2008})}\BibitemShut {NoStop}%
\bibitem [{\citenamefont {Paar}\ \emph {et~al.}(2003)\citenamefont {Paar},
  \citenamefont {Ring}, \citenamefont {Nik{\v{s}}i{\'c}},\ and\ \citenamefont
  {Vretenar}}]{PaarRingNiksicEtAl2003}%
  \BibitemOpen
  \bibfield  {author} {\bibinfo {author} {\bibfnamefont {N.}~\bibnamefont
  {Paar}}, \bibinfo {author} {\bibfnamefont {P.}~\bibnamefont {Ring}}, \bibinfo
  {author} {\bibfnamefont {T.}~\bibnamefont {Nik{\v{s}}i{\'c}}}, \ and\
  \bibinfo {author} {\bibfnamefont {D.}~\bibnamefont {Vretenar}},\ }\href@noop
  {} {\bibfield  {journal} {\bibinfo  {journal} {Physical Review C}\ }\textbf
  {\bibinfo {volume} {67}},\ \bibinfo {pages} {034312} (\bibinfo {year}
  {2003})}\BibitemShut {NoStop}%
\bibitem [{\citenamefont {Litvinova}\ \emph {et~al.}(2008)\citenamefont
  {Litvinova}, \citenamefont {Ring},\ and\ \citenamefont
  {Tselyaev}}]{LitvinovaRingTselyaev2008}%
  \BibitemOpen
  \bibfield  {author} {\bibinfo {author} {\bibfnamefont {E.}~\bibnamefont
  {Litvinova}}, \bibinfo {author} {\bibfnamefont {P.}~\bibnamefont {Ring}}, \
  and\ \bibinfo {author} {\bibfnamefont {V.}~\bibnamefont {Tselyaev}},\
  }\href@noop {} {\bibfield  {journal} {\bibinfo  {journal} {Physical Review
  C}\ }\textbf {\bibinfo {volume} {78}},\ \bibinfo {pages} {014312} (\bibinfo
  {year} {2008})}\BibitemShut {NoStop}%
\bibitem [{\citenamefont {Litvinova}\ \emph {et~al.}(2010)\citenamefont
  {Litvinova}, \citenamefont {Ring},\ and\ \citenamefont
  {Tselyaev}}]{LitvinovaRingTselyaev2010}%
  \BibitemOpen
  \bibfield  {author} {\bibinfo {author} {\bibfnamefont {E.}~\bibnamefont
  {Litvinova}}, \bibinfo {author} {\bibfnamefont {P.}~\bibnamefont {Ring}}, \
  and\ \bibinfo {author} {\bibfnamefont {V.}~\bibnamefont {Tselyaev}},\
  }\href@noop {} {\bibfield  {journal} {\bibinfo  {journal} {Physical Review
  Letters}\ }\textbf {\bibinfo {volume} {105}},\ \bibinfo {pages} {022502}
  (\bibinfo {year} {2010})}\BibitemShut {NoStop}%
\bibitem [{\citenamefont {Litvinova}\ \emph {et~al.}(2013)\citenamefont
  {Litvinova}, \citenamefont {Ring},\ and\ \citenamefont
  {Tselyaev}}]{LitvinovaRingTselyaev2013}%
  \BibitemOpen
  \bibfield  {author} {\bibinfo {author} {\bibfnamefont {E.}~\bibnamefont
  {Litvinova}}, \bibinfo {author} {\bibfnamefont {P.}~\bibnamefont {Ring}}, \
  and\ \bibinfo {author} {\bibfnamefont {V.}~\bibnamefont {Tselyaev}},\
  }\href@noop {} {\bibfield  {journal} {\bibinfo  {journal} {Physical Review
  C}\ }\textbf {\bibinfo {volume} {88}},\ \bibinfo {pages} {044320} (\bibinfo
  {year} {2013})}\BibitemShut {NoStop}%
\bibitem [{\citenamefont {Litvinova}(2015)}]{Litvinova2015}%
  \BibitemOpen
  \bibfield  {author} {\bibinfo {author} {\bibfnamefont {E.}~\bibnamefont
  {Litvinova}},\ }\href@noop {} {\bibfield  {journal} {\bibinfo  {journal}
  {Physical Review C}\ }\textbf {\bibinfo {volume} {91}},\ \bibinfo {pages}
  {034332} (\bibinfo {year} {2015})}\BibitemShut {NoStop}%
\bibitem [{\citenamefont {Litvinova}\ \emph
  {et~al.}(2009{\natexlab{a}})\citenamefont {Litvinova}, \citenamefont {Ring},
  \citenamefont {Tselyaev},\ and\ \citenamefont
  {Langanke}}]{LitvinovaRingTselyaevEtAl2009}%
  \BibitemOpen
  \bibfield  {author} {\bibinfo {author} {\bibfnamefont {E.}~\bibnamefont
  {Litvinova}}, \bibinfo {author} {\bibfnamefont {P.}~\bibnamefont {Ring}},
  \bibinfo {author} {\bibfnamefont {V.}~\bibnamefont {Tselyaev}}, \ and\
  \bibinfo {author} {\bibfnamefont {K.}~\bibnamefont {Langanke}},\ }\href
  {\doibase 10.1103/PhysRevC.79.054312} {\bibfield  {journal} {\bibinfo
  {journal} {Physical Review C}\ }\textbf {\bibinfo {volume} {79}},\ \bibinfo
  {pages} {054312} (\bibinfo {year} {2009}{\natexlab{a}})}\BibitemShut
  {NoStop}%
\bibitem [{\citenamefont {Massarczyk}\ \emph {et~al.}(2012)\citenamefont
  {Massarczyk}, \citenamefont {Schwengner}, \citenamefont {D{\"o}nau},
  \citenamefont {Litvinova}, \citenamefont {Rusev}, \citenamefont {Beyer},
  \citenamefont {Hannaske}, \citenamefont {Junghans}, \citenamefont {Kempe},
  \citenamefont {Kelley} \emph {et~al.}}]{MassarczykSchwengnerDoenauEtAl2012}%
  \BibitemOpen
  \bibfield  {author} {\bibinfo {author} {\bibfnamefont {R.}~\bibnamefont
  {Massarczyk}}, \bibinfo {author} {\bibfnamefont {R.}~\bibnamefont
  {Schwengner}}, \bibinfo {author} {\bibfnamefont {F.}~\bibnamefont
  {D{\"o}nau}}, \bibinfo {author} {\bibfnamefont {E.}~\bibnamefont
  {Litvinova}}, \bibinfo {author} {\bibfnamefont {G.}~\bibnamefont {Rusev}},
  \bibinfo {author} {\bibfnamefont {R.}~\bibnamefont {Beyer}}, \bibinfo
  {author} {\bibfnamefont {R.}~\bibnamefont {Hannaske}}, \bibinfo {author}
  {\bibfnamefont {A.}~\bibnamefont {Junghans}}, \bibinfo {author}
  {\bibfnamefont {M.}~\bibnamefont {Kempe}}, \bibinfo {author} {\bibfnamefont
  {J.~H.}\ \bibnamefont {Kelley}},  \emph {et~al.},\ }\href@noop {} {\bibfield
  {journal} {\bibinfo  {journal} {Physical Review C}\ }\textbf {\bibinfo
  {volume} {86}},\ \bibinfo {pages} {014319} (\bibinfo {year}
  {2012})}\BibitemShut {NoStop}%
\bibitem [{\citenamefont {Egorova}\ and\ \citenamefont
  {Litvinova}(2016)}]{EgorovaLitvinova2016}%
  \BibitemOpen
  \bibfield  {author} {\bibinfo {author} {\bibfnamefont {I.~A.}\ \bibnamefont
  {Egorova}}\ and\ \bibinfo {author} {\bibfnamefont {E.}~\bibnamefont
  {Litvinova}},\ }\href@noop {} {\bibfield  {journal} {\bibinfo  {journal}
  {Physical Review C}\ }\textbf {\bibinfo {volume} {94}},\ \bibinfo {pages}
  {034322} (\bibinfo {year} {2016})}\BibitemShut {NoStop}%
\bibitem [{\citenamefont {Endres}\ \emph {et~al.}(2010)\citenamefont {Endres},
  \citenamefont {Litvinova}, \citenamefont {Savran}, \citenamefont {Butler},
  \citenamefont {Harakeh}, \citenamefont {Harissopulos}, \citenamefont
  {Herzberg}, \citenamefont {Kr\"ucken}, \citenamefont {Lagoyannis},
  \citenamefont {Pietralla}, \citenamefont {Ponomarev}, \citenamefont
  {Popescu}, \citenamefont {Ring}, \citenamefont {Scheck}, \citenamefont
  {Sonnabend}, \citenamefont {Stoica}, \citenamefont {W\"ortche},\ and\
  \citenamefont {Zilges}}]{EndresLitvinovaSavranEtAl2010}%
  \BibitemOpen
  \bibfield  {author} {\bibinfo {author} {\bibfnamefont {J.}~\bibnamefont
  {Endres}}, \bibinfo {author} {\bibfnamefont {E.}~\bibnamefont {Litvinova}},
  \bibinfo {author} {\bibfnamefont {D.}~\bibnamefont {Savran}}, \bibinfo
  {author} {\bibfnamefont {P.~A.}\ \bibnamefont {Butler}}, \bibinfo {author}
  {\bibfnamefont {M.~N.}\ \bibnamefont {Harakeh}}, \bibinfo {author}
  {\bibfnamefont {S.}~\bibnamefont {Harissopulos}}, \bibinfo {author}
  {\bibfnamefont {R.-D.}\ \bibnamefont {Herzberg}}, \bibinfo {author}
  {\bibfnamefont {R.}~\bibnamefont {Kr\"ucken}}, \bibinfo {author}
  {\bibfnamefont {A.}~\bibnamefont {Lagoyannis}}, \bibinfo {author}
  {\bibfnamefont {N.}~\bibnamefont {Pietralla}}, \bibinfo {author}
  {\bibfnamefont {V.~Y.}\ \bibnamefont {Ponomarev}}, \bibinfo {author}
  {\bibfnamefont {L.}~\bibnamefont {Popescu}}, \bibinfo {author} {\bibfnamefont
  {P.}~\bibnamefont {Ring}}, \bibinfo {author} {\bibfnamefont {M.}~\bibnamefont
  {Scheck}}, \bibinfo {author} {\bibfnamefont {K.}~\bibnamefont {Sonnabend}},
  \bibinfo {author} {\bibfnamefont {V.~I.}\ \bibnamefont {Stoica}}, \bibinfo
  {author} {\bibfnamefont {H.~J.}\ \bibnamefont {W\"ortche}}, \ and\ \bibinfo
  {author} {\bibfnamefont {A.}~\bibnamefont {Zilges}},\ }\href {\doibase
  10.1103/PhysRevLett.105.212503} {\bibfield  {journal} {\bibinfo  {journal}
  {Physical Review Letters}\ }\textbf {\bibinfo {volume} {105}},\ \bibinfo
  {pages} {212503} (\bibinfo {year} {2010})}\BibitemShut {NoStop}%
\bibitem [{\citenamefont {Endres}\ \emph {et~al.}(2012)\citenamefont {Endres},
  \citenamefont {Savran}, \citenamefont {Butler}, \citenamefont {Harakeh},
  \citenamefont {Harissopulos}, \citenamefont {Herzberg}, \citenamefont
  {Kr\"ucken}, \citenamefont {Lagoyannis}, \citenamefont {Litvinova},
  \citenamefont {Pietralla}, \citenamefont {Ponomarev}, \citenamefont
  {Popescu}, \citenamefont {Ring}, \citenamefont {Scheck}, \citenamefont
  {Schl\"uter}, \citenamefont {Sonnabend}, \citenamefont {Stoica},
  \citenamefont {W\"ortche},\ and\ \citenamefont
  {Zilges}}]{EndresSavranButlerEtAl2012}%
  \BibitemOpen
  \bibfield  {author} {\bibinfo {author} {\bibfnamefont {J.}~\bibnamefont
  {Endres}}, \bibinfo {author} {\bibfnamefont {D.}~\bibnamefont {Savran}},
  \bibinfo {author} {\bibfnamefont {P.~A.}\ \bibnamefont {Butler}}, \bibinfo
  {author} {\bibfnamefont {M.~N.}\ \bibnamefont {Harakeh}}, \bibinfo {author}
  {\bibfnamefont {S.}~\bibnamefont {Harissopulos}}, \bibinfo {author}
  {\bibfnamefont {R.-D.}\ \bibnamefont {Herzberg}}, \bibinfo {author}
  {\bibfnamefont {R.}~\bibnamefont {Kr\"ucken}}, \bibinfo {author}
  {\bibfnamefont {A.}~\bibnamefont {Lagoyannis}}, \bibinfo {author}
  {\bibfnamefont {E.}~\bibnamefont {Litvinova}}, \bibinfo {author}
  {\bibfnamefont {N.}~\bibnamefont {Pietralla}}, \bibinfo {author}
  {\bibfnamefont {V.}~\bibnamefont {Ponomarev}}, \bibinfo {author}
  {\bibfnamefont {L.}~\bibnamefont {Popescu}}, \bibinfo {author} {\bibfnamefont
  {P.}~\bibnamefont {Ring}}, \bibinfo {author} {\bibfnamefont {M.}~\bibnamefont
  {Scheck}}, \bibinfo {author} {\bibfnamefont {F.}~\bibnamefont {Schl\"uter}},
  \bibinfo {author} {\bibfnamefont {K.}~\bibnamefont {Sonnabend}}, \bibinfo
  {author} {\bibfnamefont {V.~I.}\ \bibnamefont {Stoica}}, \bibinfo {author}
  {\bibfnamefont {H.~J.}\ \bibnamefont {W\"ortche}}, \ and\ \bibinfo {author}
  {\bibfnamefont {A.}~\bibnamefont {Zilges}},\ }\href {\doibase
  10.1103/PhysRevC.85.064331} {\bibfield  {journal} {\bibinfo  {journal}
  {Physical Review C}\ }\textbf {\bibinfo {volume} {85}},\ \bibinfo {pages}
  {064331} (\bibinfo {year} {2012})}\BibitemShut {NoStop}%
\bibitem [{\citenamefont {Lanza}\ \emph {et~al.}(2014)\citenamefont {Lanza},
  \citenamefont {Vitturi}, \citenamefont {Litvinova},\ and\ \citenamefont
  {Savran}}]{LanzaVitturiLitvinovaEtAl2014}%
  \BibitemOpen
  \bibfield  {author} {\bibinfo {author} {\bibfnamefont {E.}~\bibnamefont
  {Lanza}}, \bibinfo {author} {\bibfnamefont {A.}~\bibnamefont {Vitturi}},
  \bibinfo {author} {\bibfnamefont {E.}~\bibnamefont {Litvinova}}, \ and\
  \bibinfo {author} {\bibfnamefont {D.}~\bibnamefont {Savran}},\ }\href@noop {}
  {\bibfield  {journal} {\bibinfo  {journal} {Physical Review C}\ }\textbf
  {\bibinfo {volume} {89}},\ \bibinfo {pages} {041601} (\bibinfo {year}
  {2014})}\BibitemShut {NoStop}%
\bibitem [{\citenamefont {Pellegri}\ \emph {et~al.}(2014)\citenamefont
  {Pellegri}, \citenamefont {Bracco}, \citenamefont {Crespi}, \citenamefont
  {Leoni}, \citenamefont {Camera}, \citenamefont {Lanza}, \citenamefont
  {Kmiecik}, \citenamefont {Maj}, \citenamefont {Avigo}, \citenamefont
  {Benzoni}, \citenamefont {Blasi}, \citenamefont {Boiano}, \citenamefont
  {Bottoni}, \citenamefont {Brambilla}, \citenamefont {Ceruti}, \citenamefont
  {Giaz}, \citenamefont {Million}, \citenamefont {Morales}, \citenamefont
  {Nicolini}, \citenamefont {Vandone}, \citenamefont {Wieland}, \citenamefont
  {Bazzacco}, \citenamefont {Bednarczyk}, \citenamefont {Bellato},
  \citenamefont {Birkenbach}, \citenamefont {Bortolato}, \citenamefont
  {Cederwall}, \citenamefont {Charles}, \citenamefont {Ciemala}, \citenamefont
  {Angelis}, \citenamefont {D\'esesquelles}, \citenamefont {Eberth},
  \citenamefont {Farnea}, \citenamefont {Gadea}, \citenamefont {Gernh\"auser},
  \citenamefont {G\"orgen}, \citenamefont {Gottardo}, \citenamefont {Grebosz},
  \citenamefont {Hess}, \citenamefont {Isocrate}, \citenamefont {Jolie},
  \citenamefont {Judson}, \citenamefont {Jungclaus}, \citenamefont {Karkour},
  \citenamefont {Krzysiek}, \citenamefont {Litvinova}, \citenamefont {Lunardi},
  \citenamefont {Mazurek}, \citenamefont {Mengoni}, \citenamefont
  {Michelagnoli}, \citenamefont {Menegazzo}, \citenamefont {Molini},
  \citenamefont {Napoli}, \citenamefont {Pullia}, \citenamefont {Quintana},
  \citenamefont {Recchia}, \citenamefont {Reiter}, \citenamefont {Salsac},
  \citenamefont {Siebeck}, \citenamefont {Siem}, \citenamefont {Simpson},
  \citenamefont {S\"oderstr\"om}, \citenamefont {Stezowski}, \citenamefont
  {Theisen}, \citenamefont {Ur}, \citenamefont {Dobon},\ and\ \citenamefont
  {Zieblinski}}]{PellegriBraccoCrespiEtAl2014}%
  \BibitemOpen
  \bibfield  {author} {\bibinfo {author} {\bibfnamefont {L.}~\bibnamefont
  {Pellegri}}, \bibinfo {author} {\bibfnamefont {A.}~\bibnamefont {Bracco}},
  \bibinfo {author} {\bibfnamefont {F.}~\bibnamefont {Crespi}}, \bibinfo
  {author} {\bibfnamefont {S.}~\bibnamefont {Leoni}}, \bibinfo {author}
  {\bibfnamefont {F.}~\bibnamefont {Camera}}, \bibinfo {author} {\bibfnamefont
  {E.}~\bibnamefont {Lanza}}, \bibinfo {author} {\bibfnamefont
  {M.}~\bibnamefont {Kmiecik}}, \bibinfo {author} {\bibfnamefont
  {A.}~\bibnamefont {Maj}}, \bibinfo {author} {\bibfnamefont {R.}~\bibnamefont
  {Avigo}}, \bibinfo {author} {\bibfnamefont {G.}~\bibnamefont {Benzoni}},
  \bibinfo {author} {\bibfnamefont {N.}~\bibnamefont {Blasi}}, \bibinfo
  {author} {\bibfnamefont {C.}~\bibnamefont {Boiano}}, \bibinfo {author}
  {\bibfnamefont {S.}~\bibnamefont {Bottoni}}, \bibinfo {author} {\bibfnamefont
  {S.}~\bibnamefont {Brambilla}}, \bibinfo {author} {\bibfnamefont
  {S.}~\bibnamefont {Ceruti}}, \bibinfo {author} {\bibfnamefont
  {A.}~\bibnamefont {Giaz}}, \bibinfo {author} {\bibfnamefont {B.}~\bibnamefont
  {Million}}, \bibinfo {author} {\bibfnamefont {A.}~\bibnamefont {Morales}},
  \bibinfo {author} {\bibfnamefont {R.}~\bibnamefont {Nicolini}}, \bibinfo
  {author} {\bibfnamefont {V.}~\bibnamefont {Vandone}}, \bibinfo {author}
  {\bibfnamefont {O.}~\bibnamefont {Wieland}}, \bibinfo {author} {\bibfnamefont
  {D.}~\bibnamefont {Bazzacco}}, \bibinfo {author} {\bibfnamefont
  {P.}~\bibnamefont {Bednarczyk}}, \bibinfo {author} {\bibfnamefont
  {M.}~\bibnamefont {Bellato}}, \bibinfo {author} {\bibfnamefont
  {B.}~\bibnamefont {Birkenbach}}, \bibinfo {author} {\bibfnamefont
  {D.}~\bibnamefont {Bortolato}}, \bibinfo {author} {\bibfnamefont
  {B.}~\bibnamefont {Cederwall}}, \bibinfo {author} {\bibfnamefont
  {L.}~\bibnamefont {Charles}}, \bibinfo {author} {\bibfnamefont
  {M.}~\bibnamefont {Ciemala}}, \bibinfo {author} {\bibfnamefont {G.~D.}\
  \bibnamefont {Angelis}}, \bibinfo {author} {\bibfnamefont {P.}~\bibnamefont
  {D\'esesquelles}}, \bibinfo {author} {\bibfnamefont {J.}~\bibnamefont
  {Eberth}}, \bibinfo {author} {\bibfnamefont {E.}~\bibnamefont {Farnea}},
  \bibinfo {author} {\bibfnamefont {A.}~\bibnamefont {Gadea}}, \bibinfo
  {author} {\bibfnamefont {R.}~\bibnamefont {Gernh\"auser}}, \bibinfo {author}
  {\bibfnamefont {A.}~\bibnamefont {G\"orgen}}, \bibinfo {author}
  {\bibfnamefont {A.}~\bibnamefont {Gottardo}}, \bibinfo {author}
  {\bibfnamefont {J.}~\bibnamefont {Grebosz}}, \bibinfo {author} {\bibfnamefont
  {H.}~\bibnamefont {Hess}}, \bibinfo {author} {\bibfnamefont {R.}~\bibnamefont
  {Isocrate}}, \bibinfo {author} {\bibfnamefont {J.}~\bibnamefont {Jolie}},
  \bibinfo {author} {\bibfnamefont {D.}~\bibnamefont {Judson}}, \bibinfo
  {author} {\bibfnamefont {A.}~\bibnamefont {Jungclaus}}, \bibinfo {author}
  {\bibfnamefont {N.}~\bibnamefont {Karkour}}, \bibinfo {author} {\bibfnamefont
  {M.}~\bibnamefont {Krzysiek}}, \bibinfo {author} {\bibfnamefont
  {E.}~\bibnamefont {Litvinova}}, \bibinfo {author} {\bibfnamefont
  {S.}~\bibnamefont {Lunardi}}, \bibinfo {author} {\bibfnamefont
  {K.}~\bibnamefont {Mazurek}}, \bibinfo {author} {\bibfnamefont
  {D.}~\bibnamefont {Mengoni}}, \bibinfo {author} {\bibfnamefont
  {C.}~\bibnamefont {Michelagnoli}}, \bibinfo {author} {\bibfnamefont
  {R.}~\bibnamefont {Menegazzo}}, \bibinfo {author} {\bibfnamefont
  {P.}~\bibnamefont {Molini}}, \bibinfo {author} {\bibfnamefont
  {D.}~\bibnamefont {Napoli}}, \bibinfo {author} {\bibfnamefont
  {A.}~\bibnamefont {Pullia}}, \bibinfo {author} {\bibfnamefont
  {B.}~\bibnamefont {Quintana}}, \bibinfo {author} {\bibfnamefont
  {F.}~\bibnamefont {Recchia}}, \bibinfo {author} {\bibfnamefont
  {P.}~\bibnamefont {Reiter}}, \bibinfo {author} {\bibfnamefont
  {M.}~\bibnamefont {Salsac}}, \bibinfo {author} {\bibfnamefont
  {B.}~\bibnamefont {Siebeck}}, \bibinfo {author} {\bibfnamefont
  {S.}~\bibnamefont {Siem}}, \bibinfo {author} {\bibfnamefont {J.}~\bibnamefont
  {Simpson}}, \bibinfo {author} {\bibfnamefont {P.-A.}\ \bibnamefont
  {S\"oderstr\"om}}, \bibinfo {author} {\bibfnamefont {O.}~\bibnamefont
  {Stezowski}}, \bibinfo {author} {\bibfnamefont {C.}~\bibnamefont {Theisen}},
  \bibinfo {author} {\bibfnamefont {C.}~\bibnamefont {Ur}}, \bibinfo {author}
  {\bibfnamefont {J.~V.}\ \bibnamefont {Dobon}}, \ and\ \bibinfo {author}
  {\bibfnamefont {M.}~\bibnamefont {Zieblinski}},\ }\href {\doibase
  http://dx.doi.org/10.1016/j.physletb.2014.08.029} {\bibfield  {journal}
  {\bibinfo  {journal} {Physics Letters B}\ }\textbf {\bibinfo {volume}
  {738}},\ \bibinfo {pages} {519 } (\bibinfo {year} {2014})}\BibitemShut
  {NoStop}%
\bibitem [{\citenamefont {Krzysiek}\ \emph {et~al.}(2016)\citenamefont
  {Krzysiek}, \citenamefont {Kmiecik}, \citenamefont {Maj}, \citenamefont
  {Bednarczyk}, \citenamefont {Bracco}, \citenamefont {Crespi}, \citenamefont
  {Lanza}, \citenamefont {Litvinova}, \citenamefont {Paar}, \citenamefont
  {Avigo}, \citenamefont {Bazzacco}, \citenamefont {Benzoni}, \citenamefont
  {Birkenbach}, \citenamefont {Blasi}, \citenamefont {Bottoni}, \citenamefont
  {Brambilla}, \citenamefont {Camera}, \citenamefont {Ceruti}, \citenamefont
  {Ciema\l{}a}, \citenamefont {de~Angelis}, \citenamefont {D\'esesquelles},
  \citenamefont {Eberth}, \citenamefont {Farnea}, \citenamefont {Gadea},
  \citenamefont {Giaz}, \citenamefont {G\"orgen}, \citenamefont {Gottardo},
  \citenamefont {Grebosz}, \citenamefont {Hess}, \citenamefont {Isocarte},
  \citenamefont {Jungclaus}, \citenamefont {Leoni}, \citenamefont {Ljungvall},
  \citenamefont {Lunardi}, \citenamefont {Mazurek}, \citenamefont {Menegazzo},
  \citenamefont {Mengoni}, \citenamefont {Michelagnoli}, \citenamefont
  {Milion}, \citenamefont {Morales}, \citenamefont {Napoli}, \citenamefont
  {Nicolini}, \citenamefont {Pellegri}, \citenamefont {Pullia}, \citenamefont
  {Quintana}, \citenamefont {Recchia}, \citenamefont {Reiter}, \citenamefont
  {Rosso}, \citenamefont {Salsac}, \citenamefont {Siebeck}, \citenamefont
  {Siem}, \citenamefont {S\"oderstr\"om}, \citenamefont {Ur}, \citenamefont
  {Valiente-Dobon}, \citenamefont {Wieland},\ and\ \citenamefont
  {Ziebli\'nski}}]{KrzysiekKmiecikMajEtAl2016}%
  \BibitemOpen
  \bibfield  {author} {\bibinfo {author} {\bibfnamefont {M.}~\bibnamefont
  {Krzysiek}}, \bibinfo {author} {\bibfnamefont {M.}~\bibnamefont {Kmiecik}},
  \bibinfo {author} {\bibfnamefont {A.}~\bibnamefont {Maj}}, \bibinfo {author}
  {\bibfnamefont {P.}~\bibnamefont {Bednarczyk}}, \bibinfo {author}
  {\bibfnamefont {A.}~\bibnamefont {Bracco}}, \bibinfo {author} {\bibfnamefont
  {F.~C.~L.}\ \bibnamefont {Crespi}}, \bibinfo {author} {\bibfnamefont {E.~G.}\
  \bibnamefont {Lanza}}, \bibinfo {author} {\bibfnamefont {E.}~\bibnamefont
  {Litvinova}}, \bibinfo {author} {\bibfnamefont {N.}~\bibnamefont {Paar}},
  \bibinfo {author} {\bibfnamefont {R.}~\bibnamefont {Avigo}}, \bibinfo
  {author} {\bibfnamefont {D.}~\bibnamefont {Bazzacco}}, \bibinfo {author}
  {\bibfnamefont {G.}~\bibnamefont {Benzoni}}, \bibinfo {author} {\bibfnamefont
  {B.}~\bibnamefont {Birkenbach}}, \bibinfo {author} {\bibfnamefont
  {N.}~\bibnamefont {Blasi}}, \bibinfo {author} {\bibfnamefont
  {S.}~\bibnamefont {Bottoni}}, \bibinfo {author} {\bibfnamefont
  {S.}~\bibnamefont {Brambilla}}, \bibinfo {author} {\bibfnamefont
  {F.}~\bibnamefont {Camera}}, \bibinfo {author} {\bibfnamefont
  {S.}~\bibnamefont {Ceruti}}, \bibinfo {author} {\bibfnamefont
  {M.}~\bibnamefont {Ciema\l{}a}}, \bibinfo {author} {\bibfnamefont
  {G.}~\bibnamefont {de~Angelis}}, \bibinfo {author} {\bibfnamefont
  {P.}~\bibnamefont {D\'esesquelles}}, \bibinfo {author} {\bibfnamefont
  {J.}~\bibnamefont {Eberth}}, \bibinfo {author} {\bibfnamefont
  {E.}~\bibnamefont {Farnea}}, \bibinfo {author} {\bibfnamefont
  {A.}~\bibnamefont {Gadea}}, \bibinfo {author} {\bibfnamefont
  {A.}~\bibnamefont {Giaz}}, \bibinfo {author} {\bibfnamefont {A.}~\bibnamefont
  {G\"orgen}}, \bibinfo {author} {\bibfnamefont {A.}~\bibnamefont {Gottardo}},
  \bibinfo {author} {\bibfnamefont {J.}~\bibnamefont {Grebosz}}, \bibinfo
  {author} {\bibfnamefont {H.}~\bibnamefont {Hess}}, \bibinfo {author}
  {\bibfnamefont {R.}~\bibnamefont {Isocarte}}, \bibinfo {author}
  {\bibfnamefont {A.}~\bibnamefont {Jungclaus}}, \bibinfo {author}
  {\bibfnamefont {S.}~\bibnamefont {Leoni}}, \bibinfo {author} {\bibfnamefont
  {J.}~\bibnamefont {Ljungvall}}, \bibinfo {author} {\bibfnamefont
  {S.}~\bibnamefont {Lunardi}}, \bibinfo {author} {\bibfnamefont
  {K.}~\bibnamefont {Mazurek}}, \bibinfo {author} {\bibfnamefont
  {R.}~\bibnamefont {Menegazzo}}, \bibinfo {author} {\bibfnamefont
  {D.}~\bibnamefont {Mengoni}}, \bibinfo {author} {\bibfnamefont
  {C.}~\bibnamefont {Michelagnoli}}, \bibinfo {author} {\bibfnamefont
  {B.}~\bibnamefont {Milion}}, \bibinfo {author} {\bibfnamefont {A.~I.}\
  \bibnamefont {Morales}}, \bibinfo {author} {\bibfnamefont {D.~R.}\
  \bibnamefont {Napoli}}, \bibinfo {author} {\bibfnamefont {R.}~\bibnamefont
  {Nicolini}}, \bibinfo {author} {\bibfnamefont {L.}~\bibnamefont {Pellegri}},
  \bibinfo {author} {\bibfnamefont {A.}~\bibnamefont {Pullia}}, \bibinfo
  {author} {\bibfnamefont {B.}~\bibnamefont {Quintana}}, \bibinfo {author}
  {\bibfnamefont {F.}~\bibnamefont {Recchia}}, \bibinfo {author} {\bibfnamefont
  {P.}~\bibnamefont {Reiter}}, \bibinfo {author} {\bibfnamefont
  {D.}~\bibnamefont {Rosso}}, \bibinfo {author} {\bibfnamefont {M.~D.}\
  \bibnamefont {Salsac}}, \bibinfo {author} {\bibfnamefont {B.}~\bibnamefont
  {Siebeck}}, \bibinfo {author} {\bibfnamefont {S.}~\bibnamefont {Siem}},
  \bibinfo {author} {\bibfnamefont {P.-A.}\ \bibnamefont {S\"oderstr\"om}},
  \bibinfo {author} {\bibfnamefont {C.}~\bibnamefont {Ur}}, \bibinfo {author}
  {\bibfnamefont {J.~J.}\ \bibnamefont {Valiente-Dobon}}, \bibinfo {author}
  {\bibfnamefont {O.}~\bibnamefont {Wieland}}, \ and\ \bibinfo {author}
  {\bibfnamefont {M.}~\bibnamefont {Ziebli\'nski}},\ }\href {\doibase
  10.1103/PhysRevC.93.044330} {\bibfield  {journal} {\bibinfo  {journal}
  {Physical Review C}\ }\textbf {\bibinfo {volume} {93}},\ \bibinfo {pages}
  {044330} (\bibinfo {year} {2016})}\BibitemShut {NoStop}%
\bibitem [{\citenamefont {Litvinova}\ \emph
  {et~al.}(2009{\natexlab{b}})\citenamefont {Litvinova}, \citenamefont {Loens},
  \citenamefont {Langanke}, \citenamefont {Martinez-Pinedo}, \citenamefont
  {Rauscher}, \citenamefont {Ring}, \citenamefont {Thielemann},\ and\
  \citenamefont {Tselyaev}}]{LitvinovaLoensLangankeEtAl2009}%
  \BibitemOpen
  \bibfield  {author} {\bibinfo {author} {\bibfnamefont {E.}~\bibnamefont
  {Litvinova}}, \bibinfo {author} {\bibfnamefont {H.}~\bibnamefont {Loens}},
  \bibinfo {author} {\bibfnamefont {K.}~\bibnamefont {Langanke}}, \bibinfo
  {author} {\bibfnamefont {G.}~\bibnamefont {Martinez-Pinedo}}, \bibinfo
  {author} {\bibfnamefont {T.}~\bibnamefont {Rauscher}}, \bibinfo {author}
  {\bibfnamefont {P.}~\bibnamefont {Ring}}, \bibinfo {author} {\bibfnamefont
  {F.-K.}\ \bibnamefont {Thielemann}}, \ and\ \bibinfo {author} {\bibfnamefont
  {V.}~\bibnamefont {Tselyaev}},\ }\href@noop {} {\bibfield  {journal}
  {\bibinfo  {journal} {Nuclear Physics A}\ }\textbf {\bibinfo {volume}
  {823}},\ \bibinfo {pages} {26} (\bibinfo {year}
  {2009}{\natexlab{b}})}\BibitemShut {NoStop}%
\bibitem [{\citenamefont {Schuck}\ and\ \citenamefont
  {Tohyama}(2016)}]{SchuckTohyama2016a}%
  \BibitemOpen
  \bibfield  {author} {\bibinfo {author} {\bibfnamefont {P.}~\bibnamefont
  {Schuck}}\ and\ \bibinfo {author} {\bibfnamefont {M.}~\bibnamefont
  {Tohyama}},\ }\href@noop {} {\bibfield  {journal} {\bibinfo  {journal}
  {Physical Review B}\ }\textbf {\bibinfo {volume} {93}},\ \bibinfo {pages}
  {165117} (\bibinfo {year} {2016})}\BibitemShut {NoStop}%
\bibitem [{\citenamefont {Litvinova}\ and\ \citenamefont
  {Schuck}()}]{LitvinovaSchuck2017}%
  \BibitemOpen
  \bibfield  {author} {\bibinfo {author} {\bibfnamefont {E.}~\bibnamefont
  {Litvinova}}\ and\ \bibinfo {author} {\bibfnamefont {P.}~\bibnamefont
  {Schuck}},\ }\href@noop {} {\bibinfo  {journal} {in preparation}\
  }\BibitemShut {NoStop}%
\bibitem [{\citenamefont {Dukelsky}\ \emph {et~al.}(1998)\citenamefont
  {Dukelsky}, \citenamefont {R\"opke},\ and\ \citenamefont
  {Schuck}}]{DukelskyRoepkeSchuck1998}%
  \BibitemOpen
\bibfield  {journal} {  }\bibfield  {author} {\bibinfo {author} {\bibfnamefont
  {J.}~\bibnamefont {Dukelsky}}, \bibinfo {author} {\bibfnamefont
  {G.}~\bibnamefont {R\"opke}}, \ and\ \bibinfo {author} {\bibfnamefont
  {P.}~\bibnamefont {Schuck}},\ }\href {\doibase 10.1016/S0375-9474(97)00606-4}
  {\bibfield  {journal} {\bibinfo  {journal} {Nuclear Physics}\ }\textbf
  {\bibinfo {volume} {A628}},\ \bibinfo {pages} {17} (\bibinfo {year}
  {1998})}\BibitemShut {NoStop}%
\bibitem [{\citenamefont {Tselyaev}(1989{\natexlab{a}})}]{Tselyaev1989}%
  \BibitemOpen
  \bibfield  {author} {\bibinfo {author} {\bibfnamefont {V.}~\bibnamefont
  {Tselyaev}},\ }\href@noop {} {\bibfield  {journal} {\bibinfo  {journal}
  {Soviet Journal of Nuclear Physics}\ }\textbf {\bibinfo {volume} {50}},\
  \bibinfo {pages} {780} (\bibinfo {year} {1989}{\natexlab{a}})}\BibitemShut
  {NoStop}%
\bibitem [{\citenamefont {Litvinova}\ and\ \citenamefont
  {Tselyaev}(2007)}]{LitvinovaTselyaev2007}%
  \BibitemOpen
  \bibfield  {author} {\bibinfo {author} {\bibfnamefont {E.}~\bibnamefont
  {Litvinova}}\ and\ \bibinfo {author} {\bibfnamefont {V.}~\bibnamefont
  {Tselyaev}},\ }\href@noop {} {\bibfield  {journal} {\bibinfo  {journal}
  {Physical Review C}\ }\textbf {\bibinfo {volume} {75}},\ \bibinfo {pages}
  {054318} (\bibinfo {year} {2007})}\BibitemShut {NoStop}%
\bibitem [{\citenamefont {Tselyaev}(2007)}]{Tselyaev2007}%
  \BibitemOpen
  \bibfield  {author} {\bibinfo {author} {\bibfnamefont {V.}~\bibnamefont
  {Tselyaev}},\ }\href@noop {} {\bibfield  {journal} {\bibinfo  {journal}
  {Physical Review C}\ }\textbf {\bibinfo {volume} {75}},\ \bibinfo {pages}
  {024306} (\bibinfo {year} {2007})}\BibitemShut {NoStop}%
\bibitem [{\citenamefont {Lalazissis}\ \emph {et~al.}(2009)\citenamefont
  {Lalazissis}, \citenamefont {Karatzikos}, \citenamefont {Fossion},
  \citenamefont {Arteaga}, \citenamefont {Afanasjev},\ and\ \citenamefont
  {Ring}}]{LalazissisKaratzikosFossionEtAl2009}%
  \BibitemOpen
  \bibfield  {author} {\bibinfo {author} {\bibfnamefont {G.~A.}\ \bibnamefont
  {Lalazissis}}, \bibinfo {author} {\bibfnamefont {S.}~\bibnamefont
  {Karatzikos}}, \bibinfo {author} {\bibfnamefont {R.}~\bibnamefont {Fossion}},
  \bibinfo {author} {\bibfnamefont {D.~P.}\ \bibnamefont {Arteaga}}, \bibinfo
  {author} {\bibfnamefont {A.~V.}\ \bibnamefont {Afanasjev}}, \ and\ \bibinfo
  {author} {\bibfnamefont {P.}~\bibnamefont {Ring}},\ }\href@noop {} {\bibfield
   {journal} {\bibinfo  {journal} {Physics Letters B}\ }\textbf {\bibinfo
  {volume} {671}},\ \bibinfo {pages} {36} (\bibinfo {year} {2009})}\BibitemShut
  {NoStop}%
\bibitem [{\citenamefont {Paar}\ \emph {et~al.}(2004)\citenamefont {Paar},
  \citenamefont {Nik{\v{s}}i{\'c}}, \citenamefont {Vretenar},\ and\
  \citenamefont {Ring}}]{PaarNiksicVretenarEtAl2004a}%
  \BibitemOpen
  \bibfield  {author} {\bibinfo {author} {\bibfnamefont {N.}~\bibnamefont
  {Paar}}, \bibinfo {author} {\bibfnamefont {T.}~\bibnamefont
  {Nik{\v{s}}i{\'c}}}, \bibinfo {author} {\bibfnamefont {D.}~\bibnamefont
  {Vretenar}}, \ and\ \bibinfo {author} {\bibfnamefont {P.}~\bibnamefont
  {Ring}},\ }\href {\doibase 10.1103/PhysRevC.69.054303} {\bibfield  {journal}
  {\bibinfo  {journal} {Physical Review C}\ }\textbf {\bibinfo {volume} {69}},\
  \bibinfo {pages} {054303} (\bibinfo {year} {2004})}\BibitemShut {NoStop}%
\bibitem [{\citenamefont {Serra}(2001)}]{Serra2001}%
  \BibitemOpen
  \bibfield  {author} {\bibinfo {author} {\bibfnamefont {M.}~\bibnamefont
  {Serra}},\ }\emph {\bibinfo {title} {Field Theoretical Description of Pairing
  Correlations in Nuclear Systems}},\ \href@noop {} {Ph.D. thesis},\ \bibinfo
  {school} {Technische Universit\"at M\"unchen} (\bibinfo {year}
  {2001})\BibitemShut {NoStop}%
\bibitem [{\citenamefont {Adachi}\ and\ \citenamefont
  {Schuck}(1989)}]{AdachiSchuck1989}%
  \BibitemOpen
  \bibfield  {author} {\bibinfo {author} {\bibfnamefont {S.}~\bibnamefont
  {Adachi}}\ and\ \bibinfo {author} {\bibfnamefont {P.}~\bibnamefont
  {Schuck}},\ }\href@noop {} {\bibfield  {journal} {\bibinfo  {journal}
  {Nuclear Physics A}\ }\textbf {\bibinfo {volume} {496}},\ \bibinfo {pages}
  {485} (\bibinfo {year} {1989})}\BibitemShut {NoStop}%
\bibitem [{\citenamefont {Litvinova}\ and\ \citenamefont
  {Ring}(2006)}]{LitvinovaRing2006}%
  \BibitemOpen
  \bibfield  {author} {\bibinfo {author} {\bibfnamefont {E.}~\bibnamefont
  {Litvinova}}\ and\ \bibinfo {author} {\bibfnamefont {P.}~\bibnamefont
  {Ring}},\ }\href@noop {} {\bibfield  {journal} {\bibinfo  {journal} {Physical
  Review C}\ }\textbf {\bibinfo {volume} {73}},\ \bibinfo {pages} {044328}
  (\bibinfo {year} {2006})}\BibitemShut {NoStop}%
\bibitem [{\citenamefont {Litvinova}(2012)}]{Litvinova2012}%
  \BibitemOpen
  \bibfield  {author} {\bibinfo {author} {\bibfnamefont {E.}~\bibnamefont
  {Litvinova}},\ }\href@noop {} {\bibfield  {journal} {\bibinfo  {journal}
  {Physical Review C}\ }\textbf {\bibinfo {volume} {85}},\ \bibinfo {pages}
  {021303} (\bibinfo {year} {2012})}\BibitemShut {NoStop}%
\bibitem [{\citenamefont {Tselyaev}(1989{\natexlab{b}})}]{Tselyaev1989a}%
  \BibitemOpen
  \bibfield  {author} {\bibinfo {author} {\bibfnamefont {V.}~\bibnamefont
  {Tselyaev}},\ }\href@noop {} {\bibfield  {journal} {\bibinfo  {journal}
  {Soviet Journal of Nuclear Physics}\ }\textbf {\bibinfo {volume} {50}},\
  \bibinfo {pages} {1252} (\bibinfo {year} {1989}{\natexlab{b}})}\BibitemShut
  {NoStop}%
\bibitem [{\citenamefont {Kamerdzhiev}\ \emph {et~al.}(2004)\citenamefont
  {Kamerdzhiev}, \citenamefont {Speth},\ and\ \citenamefont
  {Tertychny}}]{KamerdzhievSpethTertychny2004}%
  \BibitemOpen
  \bibfield  {author} {\bibinfo {author} {\bibfnamefont {S.}~\bibnamefont
  {Kamerdzhiev}}, \bibinfo {author} {\bibfnamefont {J.}~\bibnamefont {Speth}},
  \ and\ \bibinfo {author} {\bibfnamefont {G.}~\bibnamefont {Tertychny}},\
  }\href {\doibase http://dx.doi.org/10.1016/j.physrep.2003.11.001} {\bibfield
  {journal} {\bibinfo  {journal} {Physics Reports}\ }\textbf {\bibinfo {volume}
  {393}},\ \bibinfo {pages} {1 } (\bibinfo {year} {2004})}\BibitemShut
  {NoStop}%
\bibitem [{\citenamefont {Tselyaev}\ \emph {et~al.}(2007)\citenamefont
  {Tselyaev}, \citenamefont {Speth}, \citenamefont {Grummer}, \citenamefont
  {Krewald}, \citenamefont {Avdeeenkov}, \citenamefont {Litvinova},\ and\
  \citenamefont {Tertychny}}]{TselyaevSpethGrummerEtAl2007}%
  \BibitemOpen
  \bibfield  {author} {\bibinfo {author} {\bibfnamefont {V.}~\bibnamefont
  {Tselyaev}}, \bibinfo {author} {\bibfnamefont {J.}~\bibnamefont {Speth}},
  \bibinfo {author} {\bibfnamefont {F.}~\bibnamefont {Grummer}}, \bibinfo
  {author} {\bibfnamefont {S.}~\bibnamefont {Krewald}}, \bibinfo {author}
  {\bibfnamefont {A.}~\bibnamefont {Avdeeenkov}}, \bibinfo {author}
  {\bibfnamefont {E.}~\bibnamefont {Litvinova}}, \ and\ \bibinfo {author}
  {\bibfnamefont {G.}~\bibnamefont {Tertychny}},\ }\href@noop {} {\bibfield
  {journal} {\bibinfo  {journal} {Physical Review C}\ }\textbf {\bibinfo
  {volume} {75}},\ \bibinfo {pages} {014315} (\bibinfo {year}
  {2007})}\BibitemShut {NoStop}%
\bibitem [{\citenamefont {Ring}\ \emph {et~al.}(2001)\citenamefont {Ring},
  \citenamefont {Ma}, \citenamefont {Van~Giai}, \citenamefont {Vretenar},
  \citenamefont {Wandelt},\ and\ \citenamefont {Cao}}]{RingMaVanGiaiEtAl2001}%
  \BibitemOpen
  \bibfield  {author} {\bibinfo {author} {\bibfnamefont {P.}~\bibnamefont
  {Ring}}, \bibinfo {author} {\bibfnamefont {Z.-Y.}\ \bibnamefont {Ma}},
  \bibinfo {author} {\bibfnamefont {N.}~\bibnamefont {Van~Giai}}, \bibinfo
  {author} {\bibfnamefont {D.}~\bibnamefont {Vretenar}}, \bibinfo {author}
  {\bibfnamefont {A.}~\bibnamefont {Wandelt}}, \ and\ \bibinfo {author}
  {\bibfnamefont {L.-G.}\ \bibnamefont {Cao}},\ }\href@noop {} {\bibfield
  {journal} {\bibinfo  {journal} {Nuclear Physics A}\ }\textbf {\bibinfo
  {volume} {694}},\ \bibinfo {pages} {249} (\bibinfo {year}
  {2001})}\BibitemShut {NoStop}%
\bibitem [{\citenamefont {Tselyaev}(2013)}]{Tselyaev2013}%
  \BibitemOpen
  \bibfield  {author} {\bibinfo {author} {\bibfnamefont {V.~I.}\ \bibnamefont
  {Tselyaev}},\ }\href@noop {} {\bibfield  {journal} {\bibinfo  {journal}
  {Physical Review C}\ }\textbf {\bibinfo {volume} {88}},\ \bibinfo {pages}
  {054301} (\bibinfo {year} {2013})}\BibitemShut {NoStop}%
\bibitem [{\citenamefont {Audi}\ \emph {et~al.}(2003)\citenamefont {Audi},
  \citenamefont {Wapstra},\ and\ \citenamefont
  {Thibault}}]{AudiWapstraThibault2003}%
  \BibitemOpen
  \bibfield  {author} {\bibinfo {author} {\bibfnamefont {G.}~\bibnamefont
  {Audi}}, \bibinfo {author} {\bibfnamefont {A.}~\bibnamefont {Wapstra}}, \
  and\ \bibinfo {author} {\bibfnamefont {C.}~\bibnamefont {Thibault}},\
  }\href@noop {} {\bibfield  {journal} {\bibinfo  {journal} {Nuclear Physics
  A}\ }\textbf {\bibinfo {volume} {729}},\ \bibinfo {pages} {337} (\bibinfo
  {year} {2003})}\BibitemShut {NoStop}%
\bibitem [{a()}]{a}%
  \BibitemOpen
  \href@noop {} {\emph {\bibinfo {title} {ENSDF Database,
  www.nndc.bnl.gov/ensdf/}}}\BibitemShut {NoStop}%
\bibitem [{\citenamefont {Dickhoff}\ and\ \citenamefont
  {Neck}(2005)}]{Dickhoff2005}%
  \BibitemOpen
  \bibfield  {author} {\bibinfo {author} {\bibfnamefont {W.~H.}\ \bibnamefont
  {Dickhoff}}\ and\ \bibinfo {author} {\bibfnamefont {D.~V.}\ \bibnamefont
  {Neck}},\ }\href@noop {} {\emph {\bibinfo {title} {Many-Body Theory
  Exposed!}}}\ (\bibinfo  {publisher} {World Scientific},\ \bibinfo {year}
  {2005})\BibitemShut {NoStop}%
\end{thebibliography}%
\end{document}